\documentclass[12pt]{article}
\usepackage{amsfonts}
\usepackage{amsmath,amssymb}
\usepackage{graphicx}
\usepackage{cite}

\newcommand{\braket}[1]{\langle{#1}\rangle}

\newcommand{\degK}{^\circ\mathrm{K}~}
\newcommand{\mbo}[1]{$#1$}
\newcommand{\MPl}{M_{\rm Pl}}
\newcommand{\mpl}{M_{\rm Pl}}
\newcommand{\aPl}{a_{\rm Pl}}
\newcommand{\power}[1]{\times 10^{#1}}
\newcommand{\crn}{\nn \\ }
\newcommand{\bea}{\begin{eqnarray}}
\newcommand{\eea}{\end{eqnarray}}
\newcommand{\ba}{\begin{eqnarray*}}
\newcommand{\ea}{\end{eqnarray*}}

\newcommand{\nn}{\nonumber}

\newcommand{\eps}{\varepsilon}

\newcommand{\gv}{\mbox{GeV}}

\newcommand{\epo}{\,.}
\newcommand{\eco}{\,,}
\newcommand{\semis}{\,;\;\;}

\newcommand{\be}{\begin{equation}}
\newcommand{\ee}{\end{equation}}

\newcommand{\MS}{MS }
\newcommand{\MSb}{$\overline{\rm MS}$ }

\newcommand{\MSbm}{\overline{\rm MS}}
\newcommand{\veps}{\varepsilon}
\newcommand{\cL}{{\cal L}}
\newcommand{\sign}{\mbox{sign}}
\newcommand{\Reg}{\mbox{Reg}}
\newcommand{\D}{\mathrm{d}}
\newcommand{\E}{\mathrm{e}}
\newcommand{\I}{\mathrm{i}}
\newcommand{\cC}{{\cal C}}

\newcommand{\lpl}{\Lambda_{\rm Pl}}

\newcommand{\gapprox}{\raisebox{-.2ex}{$\stackrel{\textstyle>}{\raisebox{-.6ex}[0ex][0ex]{$\sim$}}$}}

\newcommand{\npb}{{\em Nucl.\ Phys.\ }{\bf B}}
\newcommand{\np}{{\em Nucl.\ Phys.\ }{\bf B}}
\newcommand{\pl}{{\em Phys.\ Lett.\ }}

\newcommand{\prl}{{\em Phys.\ Rev.\ Lett.\ } }
\newcommand{\pr}{{\em Phys.\ Rev.\ }}

\begin{document}
\title{
\vskip-3cm{\baselineskip14pt
\centerline{\normalsize DESY~13-074,~~HU-EP-13/23\hfill}
\centerline{\normalsize April 2013 / update June 2014\hfill}}
\vskip1.5cm
The Standard model as a low-energy effective theory: what is triggering
the Higgs mechanism?\footnote{I dedicate this article to the memory of
my longtime friend and colleague, Prof.~Dr.~Jochem Fleischer, who
recently passed away. The one-loop on-shell versus \MSb matching
conditions used in the present work we have worked out together more than
30 years ago.}
}

\author{
{\sc Fred Jegerlehner},
\\
\\
{\normalsize Humboldt-Universit\"at zu Berlin, Institut f\"ur Physik,}\\
{\normalsize  Newtonstrasse 15, D-12489 Berlin, Germany}\\
{\normalsize Deutsches Elektronen-Synchrotron (DESY),}\\
{\normalsize Platanenallee 6, D-15738 Zeuthen, Germany}
}

\date{}

\maketitle
\abstract{
The discovery of the Higgs by ATLAS and CMS at the LHC not only
provided the last missing building block of the electroweak Standard
Model, the mass of the Higgs has been found to have a very peculiar
value, about 126 GeV, which is such that vacuum stability may be
extending up to the Planck scale. We emphasize the consequences for
the running masses and we reconsider the role of quadratic
divergences. A change of sign of the coefficient of the
quadratically divergent terms, showing up at about $\mu_0\sim 1.4
\times 10^{16}~\gv$, may be understood as a first order phase
transition restoring the symmetric phase in the early universe, while
its large negative values at lower scales triggers the Higgs
mechanism. Running parameters evolve in such a way that the symmetry
is restored two orders of magnitude below the Planck scale. As a
consequence, the electroweak phase transition takes place near the
scale $\mu_0$ much closer to the Planck scale than anticipated so
far. The SM Higgs system and its phase transition plays a key
role for the inflation of the early universe.  Dark energy triggering
inflation is provided by the huge bare Higgs mass term and a Higgs
induced vacuum density in the symmetric phase at times before the
electroweak phase transition takes place.

\medskip

\noindent
PACS numbers: 14.80.Bn,\,11.10.Gh,\,12.15.Lk,\,98.80.Cq\\
Keywords: Vacuum stability, Higgs mechanism, inflation.
}

\newpage

\section{Introduction}
Evidence strengthens more and more that the new particle discovered by
ATLAS~\cite{ATLAS} and CMS~\cite{CMS} at the LHC at CERN is the last
missing state required by the Standard Model (SM) of particle physics~\cite{SM,QCD},
the Higgs boson~\cite{Higgs}. For the first time the complete SM spectrum is known
now. With its discovery, the mass of the Higgs boson has been
established within a narrow range such that all SM parameters (except
for the neutrino ones) for the first time are known with remarkable
accuracy. One of the interesting consequences is that now we can
answer quite reliably the long standing question where the effective SM parameters
evolve when going to highest energies.
It may be no accident that the observed Higgs mass turned out
to match expectations from considerations of SM Higgs vacuum stability
bounds, addressed long ago in Ref.~\cite{Hambye:1996wb}, for example,
and more recently in
Refs.~\cite{Holthausen:2011aa,Yukawa:3,degrassi,Moch12,Mihaila:2012fm,Chetyrkin:2012rz,Masina:2012tz,Bednyakov:2012rb,Bednyakov:2012en,Chetyrkin:2013wya,Bednyakov:2013eba,Buttazzo:2013uya,Bednyakov:2013cpa}.

Knowing the Higgs mass allows us to say more about the phase structure
of the SM. Commonly, quadratic divergences are considered to bring the
SM into trouble: the hierarchy, fine tuning or naturalness problem. If
one understands the SM as the renormalizable~\cite{tHooft71} low-energy effective tail
of a system existing at the Planck scale, which exhibits the inverse
Planck length $\Lambda_{\rm Pl}= (G)^{-1/2}\simeq 1.22\power{19}~\gv$
($G$ Newton's gravitational constant) as a fundamental
cutoff\footnote{We will used Planck mass and Planck cutoff synonymous
i.e.  $M_\mathrm{Pl}=\Lambda_\mathrm{Pl}\approx 10^{19}~\gv$.}, the
relation between bare and renormalized parameters acquires a direct
physical meaning. The low-energy expansion in the small parameter
$x=E/\Lambda_{\rm Planck}$ suggests that only operators of dimension 4
or less are seen at low energies, which means that the low-energy tail
is a local renormalizable effective Lagrangian Quantum Field Theory
(QFT). As energies increase at some point the first non-renormalizable
effective interactions show up: operators of dimension 5 involving
fermion fields and operators of dimension 6 which can be build from
bosonic fields or by four-fermion structures. Dimension 5 operators
yield typically a 0.1\% effect, a typical accuracy achieved in many
particle physics experiments, at what is typical for Grand Unified
Theory (GUT) scales, namely $10^{16}~\gv$. Dimension 6 operators would
yield an effect of similar size at $\sim 3.6 \times 10^{18}~\gv$. So
we can expect local renormalizable QFT structure to apply up to about
two orders of magnitude below the Planck scale, because the infinite
tower of non-renormalizable operators scaling like $x^n$ with
$n=1,2,3,\cdots$ are irrelevant, i.e. they scale down with increasing
powers of the inverse cutoff. The trouble makers are the relevant
operators, those which have positive mass dimension: the mass terms in
particular. The latter scale like $\Lambda_{\rm Planck}/m_f$ for
fermions and like $\Lambda_{\rm Planck}^2/M^2_b$ for bosons. As
relevant operators they have to be tuned in order not to freeze out by
acquiring effective masses scaled up by one or two power in the cutoff
for fermions or bosons, respectively. In condensed matter physics one
would tune, as a typical relevant parameter, the temperature $T$ to
its critical value $T_c$, in order to let the system built up long
range correlations, known as critical phenomena (see
e.g. Ref.~\cite{Jegerlehner:1974dd}).  In particle physics the role
of the reduced temperature $(T-T_c)/T_c$ is taken by the renormalized
particle mass, which has to remain small enough in order the particle
is seen in the low-energy spectrum. What is tuning particle masses in
the low-energy effective theory? Symmetries as we know!  Chiral
symmetry protects the fermion masses, local gauge symmetry protects
the gauge boson masses, their non-vanishing being a consequence of
spontaneous symmetry breaking. The one exception are scalar masses,
which only can be protected by doubling the states by pairing all SM
particles, supplemented by an additional Higgs doublet, into a
supersymmetric extension. Alternatively, a conformal conspiracy could
be at work when the entire particle content of the SM or an extension
of it is such that the fermionic and bosonic degrees of freedom
compensate each other collectively. The well known example is
Veltman's ``Infrared - Ultraviolet Connection'' proposed in
Ref.~\cite{Veltman:1980mj} (see also
Refs.~\cite{Decker:1979cw,Degrassi:1992ff,Fang:1996cn}), which noted
that the coefficient of the quadratic divergences could vanish if the
sum of properly weighted Higgs, $W$ and $Z$ boson mass-squares would
cancel the top quark mass-square contribution (see below).

One of the key indications that the SM is a low-energy effective
theory is the occurrence of local gauge symmetries. In particular, the
non-Abelian local symmetries are not symmetries in the usual sense,
like global symmetries. They rather represent a dynamical principle
(like the equivalence principle in gravity) implying a special form of
the dynamics. One could call them ``quantum symmetries'' as they
determine a form of quantum interference known as gauge
cancellations. The latter are well known from processes like $W$-pair
production in $e^+e^-$-annihilation, where three Born level diagrams conspire to produce large
cancellations of terms growing badly with energy and as a result yield
the tamed observable cross section (see
e.g. Ref.~\cite{Jegerlehner:1994zp} and references therein). In fact
a non-Abelian gauge structure is an automatic consequence of a
low-energy expansion: it is the only possible residual interaction
structure, which is not suppressed by the cutoff (often misleadingly
referred to as ``tree unitarity''\footnote{Terms in tree level
amplitudes which grow faster with energy than those present in the
renormalizable spontaneously broken Yang-Mills theory are required to
be absent, since formally they seem to violate unitarity. In the low
energy expansion these terms are not absent but suppressed by large
factors $E/\Lambda$, which are not seen because the cut-off is very
large.}
constraint)~\cite{Veltman:1968ki,LlewellynSmith:1973ey,Bell:1973ex,Cornwall:1973tb,Cornwall:1974km}.
Note that spin 1 fields at long distances appear in a natural way via
multipole excitations in the Planck medium~\cite{Jegerlehner:1978nk}.
Also anomaly cancellations may be understood as low-energy
conspiracies, the otherwise non-renormalizable terms are suppressed by
inverse powers of the cutoff. The grouping of the SM fermions into
families is a consequence of this. For a more general view on the
emergence of the SM see
Refs.~\cite{Jegerlehner:1978nk,Jegerlehner:1998kt}. The general set
up for the construction of a long range effective theory is Wilson's
Renormalization Group
(RG)~\cite{Wilson:1971bg,Wilson:1971dh,Wilson:1971dc,Wilson:1973jj} of
integrating out short distance fluctuations while keeping the infrared
tail. What emerges from Wilson's RG in the infrared is equivalent to a
continuum quantum field theory RG as we know it from the SM or
elsewhere.

While the RG evolution equations in the symmetric phase of the SM have
been known for a long time to two loops, recently also the three loop
result have been calculated in
Refs.~\cite{Mihaila:2012fm,Bednyakov:2012rb,Bednyakov:2012en,Bednyakov:2013eba,Chetyrkin:2012rz,Chetyrkin:2013wya,Bednyakov:2013cpa} in the
\MSb scheme. The latter is most suitable for investigating the high-energy
behavior of the SM, which is expected to be represented by the
symmetric phase\footnote{Note that in the symmetric phase where all
field but the Higgses are massless an S-matrix does not exist, at
least in perturbation theory, and
correspondingly an on-shell scheme is not well-defined, because of the
``infrared catastrophe''. The \MSb parametrization is then a natural
parametrization at hand, most tightly related to the bare
parameters.}. The more critical point is the experimental values of
the \MSb parameters at the $Z$ boson mass or at the electroweak scale
$v=246.22~\gv$. Most parameters are known from ``low-energy''
experiments obtained in the real world in the broken phase of the SM,
typically in the on-shell renormalization scheme. The transcription of
data from the on-shell to the \MSb scheme is non-trivial within the SM
because of non-decoupling effects in the weak sector of the SM at low
energies (see
e.g. Ref.~\cite{Jegerlehner:2012kn} for a discussion in our context).

Another, maybe more serious, issue which is very different for the
electroweak (EW) sector in comparison to massless QCD, is the
appearance of quadratic divergences. They are absent in massless QCD
where in the chiral limit only logarithmic divergences show up. In the
electroweak part of the SM, by the fact that spontaneous breaking of
the symmetry does not affect the ultraviolet (UV) properties of the
theory, quadratic divergences show up in the renormalization of the
mass parameter $m^2$ of the scalar potential, which in the symmetric
phase is given by $V = \frac{m^2}{2}
\phi^2 + \frac{\lambda}{24} \phi^4 \,,$ where $\phi$ denotes the real
scalar Higgs field. The limit $m=0$ in the SM is not protected by any
symmetry, the famous naturalness or hierarchy problem. A non-zero
quadratically UV divergent $m^2$--term in the Lagrangian in any case
is induced by renormalization. Besides $m^2$, the $U(1)_Y$ and
$SU(2)_L$ gauge couplings $g'$ and $g$, respectively, the Yukawa
couplings $y_f$ and the Higgs self-coupling $\lambda$ are
logarithmically divergent only and their running is governed by the
standard RG equations for dimensionless parameters. This carries over
to the broken phase which represents the low-energy structure of the
SM. The dimensionful parameter $m^2$ transmutes to the Higgs mass
$M_H^2= \frac13\,\lambda v^2\hat{=}2m^2$ and since $\lambda$ satisfies
a normal RG equation governed by logarithmic divergences only, all
quadratic divergences must by exhibited by the Higgs bare vacuum
expectation value $v_0$, or, equivalently, in the bare Fermi constant
$G_{F0}=1/(\sqrt{2}v_0^2)$. Since all masses are proportional to $v$
all masses are affected by the issue of quadratic divergences. In the
broken phase the quadratic divergences show up in the tadpole
contributions. Renormalizability guarantees that no other type of UV
singularities are induced by renormalization, in other words a
renormalizable theory is closed with respect to dimension $d\leq 4$
operators (assuming dimensional counting within a renormalizable
gauge).

If we do not take into consideration supersymmetric extensions of the
SM, which is a possible solution of the naturalness problem, an
alternative possibility within the framework of the SM could be a
``conformal conspiracy''\footnote{As in the theory of critical
phenomena, long distance (low-energy) effective theories are
systematically constructable by applying Wilson's renormalization
group approach, and mass parameters similar to the temperature in
condensed matter physics have to be tuned to the critical surface in
parameter space~\cite{Jegerlehner:1978nk,Aoki:2012xs}. The idea is
that the statistical fluctuations at the Planck scale conspire to
select modes which are able to survive as long range correlations
(light particles). Natural are conspiracies involving few fields:
singlets, doublets, triplets as they actually appear in the
SM. Note
that GUTs are unnatural in such a low energy effective scenario, where
symmetries show up because we don't see the details of the underlying
model. GUT scenarios assume a specific large symmetry group to exist
at the high scale and that symmetries are broken spontaneously at
most. Renormalizability is imposed to hold at the high scale.}  collectively between SM particles: the quadratic divergences
can be absent if SM fermion contributions balance against the bosonic
ones~\cite{Veltman:1980mj}. Only the heavier states are relevant numerically. At
the one loop level the quadratic divergences, which in dimensional
regularization (DR) show up as poles\footnote{The massive scalar
tadpole in $D\sim 2$ is independent of $m$ given by
$A_0(m)\stackrel{D\sim 2}{=}\frac{1}{D-2}\,\frac{\mu^2}{2\pi}$ while
for $D\sim 4$ we obtain $A_0(m)=\frac{\Lambda^2}{16\pi^2}$ when
regularized with an UV cutoff $\Lambda$.} at $D=2$, are known to be
given by
\be
\delta m_H^2= \frac{\Lambda^2}{16\pi^2}\,C_1\;;\;\; C_1=\frac{6}{v^2}(M_H^2 + M_Z^2
+2 M_W^2-4 M_t^2)
\label{quadraic1}
\ee
modulo small lighter fermion contributions. The condition for the
absence of the quadratic divergences $C_1\simeq 0$ for the given top
quark mass would require a Higgs mass $M_H\simeq 314.92~\gv$ in the
one-loop approximation.  The two-loop corrections have been calculated
in Refs.~\cite{Alsarhi:1991ji,Hamada:2012bp,Jones:2013aua} with the results 
\begin{eqnarray}
C_2&=&C_1- 2 \frac{\ln (2^6/3^3)}{16\pi^2}\, [(- 36 M_t^4 + 18 M_H^2 M_t^2
+ 3 M_H^4 + 14/3 M_Z^2 M_t^2   \nonumber \\ && - 6 M_Z^2 M_H^2
- 87 M_Z^4 - 68/3 M_W^2 M_t^2
- 12 M_W^2 M_H^2 + 144 M_W^2 M_Z^2\nonumber \\
&& - 120 M_W^4)/v^4 + 32 g_3^2 M_t^2/v^2]\epo
\label{quadraic2}
\end{eqnarray}
It turns out that the two-loop correction is moderate.  If we require
$C_2 \simeq 0$ we get the solution $M_H\simeq 253.77~\gv$ closer but
still far away from its experimentally established value.  Therefore
such a possible scenario is definitely ruled out by the data, after
the Higgs mass has been determined by ATLAS and CMS.

In this paper I advocate that quadratic divergences actually could
play an important role in a different way. In fact the coefficient of
the quadratic divergence is scale dependent and exhibits a zero as
emphasized recently by Hamada, Kawai and Oda in
Ref.~\cite{Hamada:2012bp}. While Hamada, Kawai and Oda find the zero to lie
above the Planck scale, in our analysis, relying on matching
conditions for the top quark mass analyzed
in Ref.~\cite{Jegerlehner:2012kn}\footnote{Two issues which can cause
different results (different parametrizations) are the inclusion of
tadpole contributions in the EW corrections, the other the
non-decoupling of heavy particles. It should be noted that, unlike in
a calculation where we can drop tadpoles by hand, any measurement of a
physical on-shell observable automatically includes tadpole
contributions.}, we find the zero not far below the Planck
scale. The difference originates from a different estimate of
the \MSb top quark Yukawa coupling at the $Z$ mass scale, which also
implies that the Higgs potential remain stable up to the Planck
scale. For our discussion
here it is important that a zero exists below the Planck scale, where
it has a simple physical interpretation. The corresponding
change in sign seems to provide a natural explanation for the
Higgs mechanism in the SM. In the very early universe, the quadratically enhanced
bare Higgs mass term provides a large dark energy density, which
triggers inflation. In this scenario the hierarchy problem is not a
problem but the solution which explains inflation in the evolution of
the early universe as a natural phenomenon within the SM. As the
universe is cooling down the bare Higgs mass changes sign and thus
triggers the Higgs mechanism, stops inflation and the negative $m^2$
term falls into competition with the finite temperature term and
allows for the EW phase transition. In our
\textit{``Low Energy Effective SM''} (LEESM) scenario the EW phase transition
is closely correlated to the Higgs mechanism as we will see. 

In the next Section we remind the reader about the emergence of a
local renormalizable QFT in a low energy expansion from a system
exhibiting a physical UV cutoff at the microscopic level. In Section
3 we discuss the matching conditions which determine the \MSb
parameters from their physical on-shell counterparts. We emphasize
the failure of ``decoupling by hand'' prescriptions in the weak sector
of the SM.  The evolution of the SM running parameters up to the
Planck scale is presented in Section 4 for couplings, masses and the
Higgs vacuum expectation value (VEV). Section 5 is devoted to a
discussion of the scale dependence of the quadratic divergences and
the observed first order phase transition, which triggers the Higgs
mechanism. The impact of the results on inflation scenarios is briefly
addressed in Section 6. A summary and outlook follows in Section 7.  

\section{Low energy effective QFT of a cutoff system}
If we say that the SM is a low-energy effective theory we mean that
there must exist a more fundamental system exhibiting a physical
cutoff, as typical for condensed matter systems. Such a system we
expect to reside at the Planck scale, and the SM is expected to be the
renormalizable tail at long distances relative to the Planck
length. The Planck energy scale being beyond any direct experimental
access, so far we only know its long range structure and that the
underlying fundamental system must be in the universality class of the
SM. Let us be more specific and sketch the construction of a
low-energy effective QFT by looking at the cutoff version of the
Higgs system only, for simplicity:
\bea
\cL =\cL_0+\cL_{\rm int}=\frac12
\,\partial^\mu \phi(x)\,(1+\Box/\Lambda^2)\,\partial_\mu \phi(x)
-\frac12\,m_0^2\,\phi(x)^2-\frac{\lambda_0\,\Lambda^\veps}{4!}\,\phi^4(x)\epo
\nn \\
\eea
The regularization is chosen here as a Pais-Uhlenbeck higher-derivative
kinetic term~\cite{Pais:1950za}, which is equivalent to a
Pauli-Villars cutoff~\cite{Pauli:1949zm}. We are interested in the
model for $D=4$ space-time dimensions but may consider the more
general case in $D=4-\veps$ dimensions with $2\leq D
\leq 4$ in order to make comparisons with the same model in
dimensional regularization. It is characterized by its vertex
functions (connected amputated one-particle irreducible diagrams) of
$N$ scalar fields $\Gamma^{(N)}_{\Lambda,b}(p;m_0,\lambda_0)=\langle
\tilde{\phi}(p_1)\cdots\tilde{\phi}(p_{N-1})\phi(0)\rangle^{\rm prop}$
as a function of the set of independent momenta, which we denote by
$p\hat{=}\{p_i\}\,(i=1,\cdots,N-1)$.  The bare functions are related to the renormalized
ones by (for specific renormalization conditions
see Ref.~\cite{Jegerlehner:1976xd}) reparametrizing parameters and fields
\bea
\Gamma^{(N)}_{\Lambda\,r}(p;m,\lambda)=Z^{N/2}(\Lambda/m,\lambda)\,
\Gamma^{(N)}_{\Lambda\,b}(p;\Delta m_0(\Lambda,m,\lambda),\lambda_0(\Lambda/m,\lambda))\epo
\eea
They satisfy a RG equation for the response to a
change of the cutoff $\Lambda$ for fixed renormalized parameters
$\left.\Lambda \frac{\partial}{\partial
\Lambda}\,\Gamma{(N)}_{\Lambda\,b}\,\right|_{m,\lambda}$, which by applying
the chain rule of differentiation reads
\bea
&&\left(\Lambda \frac{\partial}{\partial \Lambda}+\beta_0
\frac{\partial}{\partial \lambda} -N\,\gamma_0+ \delta_0\, \Delta
m_0^2\,\frac{\partial}{\partial \Delta
m_0^2}\right)\,\Gamma^{(N)}_{\Lambda\,b}(p;m_0,\lambda_0)\nn \\ &=&
Z^{-N/2}\,\Lambda \frac{\partial}{\partial \Lambda}\,\Gamma^{(N)}_{\Lambda\,r}(p;m,\lambda)\epo
\label{LamRG}
\eea
$m^2_{0c}$ is the ``critical value'' of the bare mass for which the
renormalized mass is zero, i.e.
$\left.\Gamma^{(2)}_{\Lambda\,b}\right|_{p=0}=0$, and $\Delta
m_0^2=m_0^2-m_{0c}^2$ corresponds to the renormalized mass parameter.
Since the renormalized vertex functions have a regular limit as
$\Lambda \to \infty$, to all orders in perturbation theory the
inhomogeneous part behaves as
\bea
Z^{N/2}\,\Lambda \frac{\partial}{\partial \Lambda}\,\Gamma^{(N)}_{\Lambda\,r}
(p; m,\lambda)=O(\Lambda^{-2}(\ln \Lambda)^l)\;,
\eea
i.e., the inhomogeneous part, representing a cutoff insertion, falls
off faster than the l.h.s. of Eq.~(\ref{LamRG}) by two powers in the
cutoff for large cutoffs. This is easy to understand given the fact
that the cutoff enters $\cL$ as a term proportional to
$\Lambda^{-2}$. Beyond perturbation theory one would have to require
the condition
\bea
Z^{N/2}\,\Lambda \frac{\partial}{\partial
\Lambda}\,\Gamma^{(N)}_{\Lambda\,r}(p;m,\lambda)/\Gamma^{(N)}_{\Lambda\,b}=O(\Lambda^{-\eta})\,,
\eea
for some positive $\eta$. In addition, also all the RG equation
coefficients exist as non-trivial functions in the limit of infinite cutoff:
\bea
\lim_{\Lambda \to \infty} \alpha_0(\Lambda/m,\lambda) =\alpha(\lambda)\,,\:\alpha=\beta,\gamma,\delta\,,
\eea
for dimensions $2\leq D \leq 4$. In $D=4-\veps$ dimensions the
proper vertex-functions have a large cutoff $\Lambda$-expansion (see Ref.~\cite{Symanzik:1975rz})
\bea
\Gamma^{(N)}_{\Lambda\,b}(p;\Delta m_0,\lambda_0)=\sum_{j,k,l \geq
0} \Lambda^{-2j-\veps k}(\ln \Lambda)^l\,f^{(N)}_{jkl}(p\;\Delta m_0, \lambda_0\Lambda^\veps)\,,
\eea
and for large $\Lambda$ we obtain the \underline{preasymptote} of $\Gamma^{(N)}_{\Lambda\,b}$
\bea
\Gamma^{(N)}_{\Lambda\,{\rm as}}(p;\Delta m_0,\lambda_0)=\sum_{k,l \geq
0} \Lambda^{-\veps k}(\ln \Lambda)^l\,f^{(N)}_{0kl}(p\;\Delta m_0, \lambda_0\Lambda^\veps)\,,
\eea
collecting the leading terms and satisfies the bound
\bea
\left|\Gamma^{(N)}_{\Lambda\,b}(p;\Delta
m_0,\lambda_0)-\Gamma^{(N)}_{\Lambda\,{\rm as}}(p;\Delta m_0,\lambda_0) \right|=O(\Lambda^{-2}(\ln \Lambda^{l_x}))\epo
\eea
The index $l_x$ is bounded to all orders in the perturbation
expansion. The key point is that the still cutoff dependent preasymptote satisfies a
\underline{homogeneous} RG equation, a special property of the long range
tail of the bare theory:
\bea
&&\left(\Lambda \frac{\partial}{\partial \Lambda}+\beta_{\rm
as}(\Lambda/\Delta m_0,\lambda_0)
\frac{\partial}{\partial \lambda_0} -N\,\gamma_{\rm
as}(\Lambda/\Delta m_0,\lambda_0)\right. \nn \\ && \left. \hspace*{1cm}
+ \delta_{\rm
as}(\Lambda/\Delta m_0,\lambda_0)\, \Delta
m_0^2\,\frac{\partial}{\partial \Delta
m_0^2}\right)\,\Gamma^{(N)}_{\Lambda\,{\rm as}}(p;\Delta m_0,\lambda_0)=0\epo
\label{preasRG}
\eea
For more details see
Refs.~\cite{Jegerlehner:1976xd,Symanzik:1975rz,Brezin:1976bp}.  The
homogeneity tells us that $\Lambda$ has lost it function as a cutoff
and takes the role of a renormalization scale, i.e.,
(\ref{preasRG}) represents the response of a rescaling of the system: a change in $\Lambda$
is compensated by a finite renormalization of the fields, the
couplings and the masses\footnote{This is similar to the well known response of the on-shell renormalized
theory to a change in the mass, now considered in the continuum limit $\Lambda \to
\infty$ renormalized QFT. It is
given by the Callan-Symanzik equation~\cite{Callan:1970yg,Symanzik:1970rt}
\begin{equation*}
\left(m \frac{\partial}{\partial m}+\beta(\lambda)
\frac{\partial}{\partial \lambda} -N\,\gamma(\lambda)\right)\,\Gamma^{(N)}_{r}(p;m,\lambda)=
-m^2\,\left(2-\delta(\lambda)\right)\,\Delta_0\,\Gamma^{(N)}_{r}(p;m,\lambda)\;,
\end{equation*}
where $\Delta_0$ is the integrated mass operator insertion. For large
momenta the r.h.s. is suppressed $O(m^2\,\ln(m)^l)$ by the small mass-square $m^2 \ll
p^2$ up to logarithms, such that for large momenta asymptotically
\begin{equation*}
\left(m \frac{\partial}{\partial m}+\beta(\lambda)
\frac{\partial}{\partial \lambda} -N\,\gamma(\lambda)\right)\,\Gamma^{(N)}_{r\,as}(p;m,\lambda)=0\,.
\end{equation*}
The mass asymptotically only plays the role of a renormalization
scale, $\Gamma^{(N)}_{r\,as}(p;m,\lambda)$ are vertex functions of an
effectively massless theory. Up to appropriate finite
reparametrization and a rescaling $m=\kappa \mu$ the homogeneous CS
equations are nothing but the standard RG equation in the \MSb
scheme.}.  By a \underline{finite} renormalization we may
reparametrize the preasymptote by imposing appropriate renormalization
conditions. Then there exist a rescaling $\Lambda =\kappa \mu$ such
that we obtain the usual RG in the renormalization scale $\mu$ of a
non-trivial continuum QFT. This provides a precise interrelation
between preasymptotic and \MSb renormalized quantities, and hence
between the bare system seen from long distance and the familiar
renormalized QFT physics. Thus, what we observe as the SM is a
physical reparametrization (renormalization) of the preasymptotic bare
theory. In the language of critical phenomena the ``bare world'' at
the Planck scale has to be in the universality class of the SM. As we
only observe the tail, details of the bare world remain largely
unknown. One of the impacts of the very high Planck scale is that the
local renormalizable QFT structure of the SM is presumably valid up to
what is a typical GUT scale. It has nothing to do with grand
unification though.  This also justifies the application of the SM RG
up to high scales. The tuning ``to criticality'' of the bare mass to
the critical mass $m_{0c}$ correspond to what is known as the
hierarchy or naturalness problem in the SM. This naturalness problem
is asking for an answer to the question ``who is tuning the knob of
the thermostat to adjust the temperature to its critical value (which
is determined by the underlying atomic structure of the condensed
matter system)''. In the symmetric phase of the SM, where masses of
fermions and gauge bosons are forbidden by the known chiral and gauge
symmetries, respectively, there is only one mass, common for all four
fields in the complex Higgs doublet, to be renormalized. Here we
encounter the fine tuning relation of the form
\bea
m_0^2=m^2+\delta m^2\;;\;\; \delta m^2= \frac{\Lambda^2}{32 \pi^2}\,C
\eea
with a coefficient typically $C=O(1)$. To keep the renormalized mass
$m$ at some small value, which can be seen at low energy, $m^2_0$ has
to be adjusted to compensate the huge number $\delta m^2$ such that
about 35 digits must be adjusted in order to get the observed value
around the electroweak scale.  This is a problem only in cases where
we have to take the relation between bare and renormalized theory
serious, like in a condensed matter system or here in the LEESM
scenario. The difference is of course that in particle physics we
never will be able to directly access experimentally the bare system
sitting at the Planck scale. Furthermore, we do not know what should
be the renormalized $m^2$ in the symmetric phase where all physics is
different anyway. The hierarchy problem thus can be reformulated as
``why is $m^2$ in the symmetric phase so much larger than $M^2_H$ in
the broken phase?'' The answer is: $m^2$ is naturally large because of
the quadratic divergences, while $M^2_H=
\frac13\,\lambda\,v^2$ is small because the order parameter $v$, which
sets the scale for the low energy mass spectrum, is naturally a long
range (low-energy) quantity (similar to the magnetization in a
ferromagnetic system). What would it mean if $v=O(\mpl)$? It would
mean that spontaneous symmetry breaking would not break the symmetry
only via an asymmetric ground state, but actually break the symmetry
at the high energy scale, i.e., the symmetry would not be recovered at
high energies. This would contradict all basic knowledge about
spontaneous symmetry breaking in physical systems.

Of course, the question we would like to understand is why $v/\mpl\sim
2\power{-17}$ is that small. In a ferromagnetic system it would mean
that the magnetization $M$ in units of the lattice spacing $a$ given
by $Ma$ is very small. The magnetization is a function of the reduced
temperature $t=(T-T_c)/T_c$ and goes to zero as $t \to -0$, so to have
$Ma$ very small mean that we are close to the critical temperature
from below. The quasi-criticality is not unnatural in our context as
the system seems to be self-tuning for its emergence at long
distances. Thus, in principle having $v$ small is not necessarily a
mystery. This question in principle can be answered by simulating the
lattice SM in the unitary gauge, where $v$ is a decent $Z_2$ order
parameter (spontaneous breaking of the symmetry $H\leftrightarrow
-H$), and can by calculated by non-perturbative means. In order to
understand the $v$ vs. $\mpl$ hierarchy more quantitatively, it would
suffice to investigate this question in a QCD, top-Yukawa, Higgs
system, where couplings must be such that all three couplings remain
asymptotically free and the Higgs vacuum stays stable up to the cutoff
(for related attempts in a different direction see
e.g.~\cite{Hegde:2013mks} and references therein). In my opinion, the
misunderstanding in arguments concerning fine-tuning problems is that
a moderately large physical number is considered to be the difference
of two large uncorrelated numbers. In fact the structure of most fine
tuning problems are different: a very large number, like
$\frac{\lpl^2}{32\pi^2}$ in our case, may be multiplied
by an $O(1)$ size function which depends on some parameters and which
exhibits a zero for particular values of the parameters. The
magnetization as a function of the temperature is a typical well known
example of this, namely, at the critical point the magnetization
necessarily gets zero, and it is naturally small if we are near the
phase transition point.

In the following we consider the SM as a strictly renormalizable
theory, regularized as usual by dimensional
regularization~\cite{dimreg} in $D=4-\veps$ space-time dimensions,
such that the \MSb parametrization and the corresponding RG can be
used in the well known form Ref.~\cite{RG}.  Some care is necessary in
applying DR when dealing with the quadratic divergences as noted in
Refs.~\cite{Veltman:1980mj,Alsarhi:1991ji,Degrassi:1992ff}. For our
LEESM scenario it is the cutoff structure of the $D=4$ world which is
relevant. It should be noted that in DR as applied to $D=4$ theories
with non-trivial spin structure the latter is always taken to reside
in $D=4$ space-time (see e.g. Sect.~2.4.2 of Ref.~\cite{mybook} for a
short outline) and in this sense is a hybrid ``analytic continuation''
designed to provide a gauge-symmetry preserving regularization of the
$D=4$ dimensional gauge theory. The DR as designed in Ref.~\cite{dimreg} is
not thought to deal with the interrelation between true $D$--dimensional
theories. Thus standard DR singles out Veltman's relation
(\ref{quadraic1}) as the relevant one against others in our case.

In order to avoid misunderstandings, the \MSb scale parameter $\mu$ in
our analysis is to be interpreted as the energy scale of physical
processes taking place at that scale, in the sense we know the
effective strong coupling $\alpha_s(M^2_\tau)$ at the $\tau$ mass
scale (e.g. hadronic $\tau$-decays) and its value $\alpha_s(M^2_Z)$ at
the $Z$ boson mass scale (e.g. hadronic $Z$-decays).  What we are
interested in is how the effective theory looks like at energies
beyond present accelerator energies. The vertex functions with scaled
up momenta for fixed parameters follows from a solution of the RG
equation. To remind the reader: a vertex function of $n_B$ boson
fields and $n_\psi$ conjugate pairs of Fermi fields in the Landau
gauge satisfies the RG equation
\bea
\left\{ \mu \frac{\partial}{\partial \mu} + \beta
\frac{\partial}{\partial g} +\gamma_m m \frac{\partial}{\partial m}-n_A \gamma_A-2n_\psi
\gamma_\psi \right\}\:\Gamma\left(\left\{ p\right\}, g, m,\mu\right)=0\;,
\eea
and is a homogeneous function of canonical dimension ${{\rm dim} \Gamma}=4-n_B-3\,n_\psi$
under rescaling of all dimensionful quantities including momenta, masses
\textbf{and} the renormalization scale $\mu=\kappa \mu_0$. The RG solution then may
be written in the form
\begin{eqnarray}
\!\!\!\!\Gamma \left(\left\{ \kappa p\right\}, g, m,\mu_0\right)&=&
\kappa^{{\rm dim} \Gamma} z_B(g,\kappa)^{-n_B}\,z_\psi(g,\kappa)^{-2n_\psi}\,
\nn \\ &\times& \Gamma \left(\left\{p\right\}, g(\kappa), \frac{m(\kappa)}{\kappa},\mu_0
\right)\,.
\label{RGsol}
\end{eqnarray}
The $z$-factors include the anomalous dimensions of the
fields (for details see e.g. Sect 2.6.5 of
Ref.~\cite{mybook}). Thus, the vertex functions at higher momenta $\{\kappa
p\}$, up to an overall factor
are given by the vertex functions at the reference momenta $\{p\}$
and reference scale $\mu_0$, e.g. $\mu_0=M_Z$,
with effective coupling $g(\kappa)$ and effective mass $m(\kappa)/\kappa$.
This is the basic type of relation for a discussion of the high energy asymptotic
behavior\footnote{A very different well known role played by the \MSb
parameter $\mu$ is the following: predictions of observables
($S$-matrix elements and related cross-sections) in fixed order
perturbation expansion are renormalization scheme dependent because of
truncation errors (missing higher order contributions, which depend on
the order of the perturbative expansion and on the
parametrization chosen). In the \MSb scheme the scheme dependence
particularly is manifest in the unphysical $\mu$ dependence of the prediction of
the physical quantity which in general gets weaker the more terms
are included.}.

Let me summarize the advantage of taking serious the idea that the SM
is a low energy effective theory of a cutoff system residing at the
Planck scale. Many structural elements usually derived form
phenomenology naturally emerge in the low energy regime we are living
in. One is the simplicity of the SM as a result of our blindness to details,
which implies more symmetries. Yang-Mills structure [gauge
cancellations] with small groups: doublets, triplets besides singlets,
Lorentz invariance\footnote{It emerges in a similar way  as rotational
invariance in condensed matter systems. Take as an example the Planck medium to be
a $d$ dimensional Euclidean lattice system.  Rotational invariance is
emerging as follows: expand the hyper-cubic lattice propagator on the
Brillouin zone
$$G_0^{-1}(\vec{q})=m_0^2+4a^{-2}\,\sum\limits_{i=1}^{d}\,\sin^2\frac{aq_i}{2}\,\to\,m_0^2+q^2
+\Lambda^{-2}q^4\semis q^2=\vec{q}^{\,2}\semis \Lambda=\pi/a$$ for
small $\vec{q}$
and replace the cutoff box by a sphere of radius $\Lambda$
$$\int\limits_{-\pi/a}^{+\pi/a}\D^d q \cdots \to
\int\limits_{|\vec{q}|\leq \Lambda}\D^d q \cdots \eco $$ up to field renormalization
which does not affect the long range properties of the original
system. For $\Lambda$ large, resulting correlation functions are
identical with those of a rotational invariant Euclidean QFT with a
cutoff.}, anomaly cancellation and family structure, triviality for
space-time dimensions $D>4$ are emergent properties. D=4 is the
border case for an interacting world at long distances, extra
dimensions just trivialize by themselves and have nothing to do with
compactification etc. Last but not least, the low energy tail is a
non-trivial renormalizable QFT. The high cutoff implies the
reliability of the LEESM scenario up to close to the Planck
scale. \textbf{This scenario does not imply that no new physics is
expected to show up even at close-by or intermediate energy scales,
but we expect it to be constrained by its natural emergence in a low
energy expansion}. Remember that the hot Planck medium residing at the
Planck scale is expected to exhibit a whole spectrum of modes, a
``chaos'' so to say, from which long range properties emerge as a
self-organizing system. We should also note that emergent low energy
symmetries are all violated near the Planck scale, which could be
important for quantities like baryon of lepton number conservation. It
is unlikely that going to higher energies what we see as the SM will
not be decorated by yet unseen physics, which still would naturally
appear as a renormalizable extension of the SM. An example could be
particle quartet conspiracies forming a low energy effective $SU(4)$
in addition to the SM gauge group.

\section{Matching conditions}
When studying the scale dependence of a theory at very high energies,
where the theory is effectively massless and hence practically in the
symmetric phase, the \MSb renormalization scheme is the favorite
choice to study the scale dependence of the theory. On the other hand
the physical values of parameters are determined by physical processes
described by on-shell matrix elements and thus usually are available
in the on-shell renormalization scheme primarily. The transition from
one scheme to the other is defined by appropriate matching
conditions. For the physical masses they are given by the mass
counterterms relating the bare and the renormalized masses as
$m_{b0}^2=M_{b}^2+\delta M^2_b$ for bosons and $m_{f0}=M_{f} + \delta
M_f$ for fermions, respectively. By $m_{i0}$ we
denoted the bare, by $m_i$ the \MSb and by $M_i$ the on-shell
masses. $\Reg=\frac{2}{\varepsilon}-\gamma +
\ln 4 \pi+\ln \mu_{0}^2$ is the UV regulator term to be set equal to
$\ln \mu^2$ where $\mu_{0}$ is the bare $\mu$-parameter while $\mu$
denotes the renormalized one. The substitution defines the UV finite
\MSb parametrization. By identifying $m_b^2=M_{b}^2+\delta
M^2_b|_{\Reg =\ln \mu^2}$ and $m_{f}=M_{f} +\delta M_f|_{\Reg =\ln
\mu^2}$, respectively, we then obtain the \MSb masses in terms of the
on-shell masses. More precisely, this follows from the following
relations valid for bosons:
\bea
m^2_{b0}=M_{b}^2+\left.\delta M^2_b\right|_{\rm OS}=m_b^2+\left.\delta M^2_b\right|_{\MSbm}\,,
\eea
where
\bea
\left.\delta M^2_b\right|_{\MSbm}=\left(\left.\delta M^2_b\right|_{\rm
OS}\right)_{\rm UV \ sing}\,,
\eea
which means that only the UV singular $\Reg$ terms are
kept as \MSb counterterms. Thus
\bea
m_b^2(\mu^2)=M_{b}^2+\left.\delta M^2_b\right|_{\rm OS}-\left.\delta
M^2_b\right|_{\MSbm}=M_{b}^2+\left(\left.\delta M^2_b\right|_{\rm
OS}\right)_{\Reg =\ln \mu^2}\epo
\eea
Corresponding linear relations hold for the fermion masses. Similar
relations apply for the coupling constants $g$, $g'$, $\lambda$ and
$y_f$, which, however, usually are fixed using the mass-coupling
relations in terms of the masses and the Higgs VEV, which is
determined by the Fermi constant as $v=(\sqrt{2}G_\mu)^{-1/2}$. Here
$G_\mu$ is the muon decay constant, which represents the Fermi
constant in the on-shell scheme. The \MSb version of the Fermi
constant we denote by $G_{F}^{\overline{\rm MS}}$ or simply by
$G_{F}$. The matching condition for the Higgs VEV may be represented
in terms of the matching condition for the muon decay constant
\bea
G_{F}^{\overline{\rm MS}}(\mu^2)=G_\mu+\left(\left.\delta G_\mu \right|_{\rm OS}\right)_{\Reg =\ln \mu^2}\,,
\eea
where $\left.\frac{\delta G_\mu}{G_\mu}\right|_{\rm OS}=2\,\frac{\delta v^{-1}}{v^{-1}}\,,$
which at one-loop is given in the Appendix. For the relevant two-loop
counterterms see Ref.~\cite{Jegerlehner:2002er,Jegerlehner:2002em}. Then the \MSb top quark
Yukawa coupling is given by
\bea
y_t^{\overline{\rm MS}}(M_t^2)=
\sqrt{2}\,\frac{m_t(M_t^2)}{v^{\overline{\rm MS}}(M_t^2)}\semis v^{\overline{\rm MS}}(\mu^2)
=\left(\sqrt{2}\,G_{F}^{\overline{\rm MS}}\right)^{-1/2}(\mu^2)\,,
\eea
and the other \MSb mass-coupling relations correspondingly.

In the mass relations just presented, tadpole contributions have to be
included in order to get a gauge invariant relationship between
on-shell and \MSb masses as well as in order to preserve the UV
singularity structure and hence the RG equations. Tadpoles show up as
renormalization counterterms of the Higgs VEV $v$ and quantities which
depend on it, in particular the masses, which are generated by the
Higgs mechanism. It is important to note that measured on-shell
observables always \textbf{include} tadpole terms. Unlike in theory
experiments cannot switch off or omit subsets of diagrams. Even
measured on-shell values of dimensionless couplings are affected
by tadpoles via the on-mass-shell condition. 

The proper expressions including the relevant tadpole terms for the SM
counterterms at one-loop have been given in
Ref.~\cite{Fleischer:1980ub} and may be found in the Appendix. For
the Higgs mass such a relation has been elaborated in
Ref.~\cite{Sirlin:1985ux} as a relation between $\lambda$ and
$\lambda_{\overline{\rm MS}}$ under the proviso that
$G_{F}^{\overline{\rm MS}}=G_\mu$, which is not generally true,
because in general $G_{F}^{\overline{\rm MS}}$ is expected to be a
running parameter as well. Interpreted as a relation between $m_H$ and
$M_H$, the relation is identical to what is obtained from the relation
$m_H^2=M_H^2+\left(\left.\delta M^2_H\right|_{\rm OS}\right)_{\Reg
=\ln \mu^2}$. Note that the only information we have on $\lambda$ is
from the experimental results on $M_H$ via
$\lambda=3\sqrt{2}G_\mu\,M_H^2$. For the top quark mass the full SM
relation between the pole mass and the \MSb mass has been evaluated
recently in Ref.~\cite{Jegerlehner:2012kn}, evaluating known results
from
Refs.~\cite{Gray:1990yh,Fleischer:1998dw,Chetyrkin:1999ys,Chetyrkin:1999qi,Melnikov:2000qh,Kataev:2010zh,Hempfling:1994ar,Jegerlehner:2003py,Jegerlehner:2003sp,Jegerlehner:2004aj,martin}
(see also Refs.~\cite{Eiras:2005yt,Faisst:2004gn} and comments in
Ref.~\cite{Jegerlehner:2012kn}) in the relation
\begin{eqnarray}
M_t - m_t(\mu^2) &=&
m_t(\mu^2) \sum_{j=1} \left( \frac{\alpha_s(\mu^2)}{\pi} \right)^j\! \rho_j 
\nn \\ && + m_t(\mu^2)\! \sum_{i=1;j=0}\! \left( \frac{\alpha(\mu^2)}{\pi} \right)^i\!
\left( \frac{\alpha_s(\mu^2)}{\pi} \right)^j\! r_{ij}\,.
\label{matching}
\end{eqnarray}
There is an almost perfect cancellation between the QCD and EW
effects for the now known value of the Higgs boson mass. While
$\left[m_t(M_t^2)-M_t\right]_{\rm QCD}=-10.38~\gv$ one finds~\cite{Jegerlehner:2012kn}
$\left[m_t(M_t^2)-M_t\right]_{\rm SM}=1.14~\gv$ for $M_H=125~\gv$.

As elaborated in Ref.~\cite{Jegerlehner:2012kn} some care is required
in the evaluation of the matching conditions. It is important to
remind that the Appelquist-Carazzone theorem~\cite{Appelquist:1974tg}
does not apply to the weak sector of the SM, i.e. we cannot
parametrize and match together effective theories by switching off
fields of mass $M>\mu$ at a given scale $\mu$.  As we know the theorem
applies to QCD and QED, and in these cases provides the basis for the
``decoupling by hand'' prescription usually used in conjunction with
the \MSb parametrization, the preferred parametrization in
perturbative QCD. The non-decoupling in the weak sector of the SM is a
consequence of the mass coupling relations, which follow if the masses
are generated by the Higgs mechanism. An important question then is
what role tadpoles play in implementing the matching conditions, since
tadpoles potentially give large contributions. However, we may take
advantage of the fact that tadpole contributions drop out from
relations between physical (on-shell) parameters and
amplitudes~\cite{Taylor76,Kraus:1997bi}, while they can produce large shifts in
the relations between the ``quasi-bare'' \MSb parameters and the
on-shell ones.  As mentioned before, potentially, the Higgs VEV $v$,
which determines the Fermi constant via $G_F=\frac{1}{\sqrt{2}v^2}$,
could be particularly affected. However, we may compare the low energy
effective Fermi constant $G_F$, given by $G_\mu$, which is determined
by the muon lifetime observed in $\mu$-decay, with its ``high
energy'' variant at the $W$ boson mass scale, where it can be
identified with $\hat{G}_\mu=\frac{12\pi
\Gamma_{W\ell\nu}}{\sqrt{2}M_W^3}$ in terms of the leptonic $W$-decay
rate. The fact that $\hat{G}_\mu\approx G_\mu$ with good accuracy is
not surprising because the tadpole corrections which potentially lead
to substantial corrections are absent in relations between observable
quantities as we know. To be precise: with the PDG values
$M_W=80.385\pm0.015~\gv$, $\Gamma_W=2.085\pm0.042~\gv$ and the
leptonic branching fraction $B(W\to\ell\nu_\ell)=10.80\pm0.09\%$ we
obtain $\hat{G}_\mu=1.15564(55)\power{-5}~\gv^{-2}$ while
$G_\mu=1.16637(1)\power{-5}~\gv^{-2}$, i.e., the on-shell Fermi constant
at scale $M_Z$ appears reduced by 0.92\% relative to $G_\mu$.

Therefore, a SM parametrization in terms of
$\alpha(M_Z),\alpha_s(M_Z), \hat{G}_\mu$ and $M_Z$ (besides the other
masses), provides a good parametrization of the observables extracted
from experiments at the vector boson mass scale. 
The $Z$ mass scale, thus is an ideal
matching scale, to evaluate the \MSb parameters in terms of
corresponding on-shell values.  
Note that the running of $G_F$ starts to be important once $M_W$, $M_Z$, $M_H$ and $M_t$ come into
play. At higher scales, certainly, the \MSb version of $v(\mu^2)$ or
equivalently $G_F^{\MSbm}(\mu^2)$ must be running as required by the
corresponding RG.

For numerical results presented in the following sections we
use values for the input parameters~\cite{pdg}:
{\small
\begin{eqnarray}
&&
M_Z = 91.1876(21)~\gv,
\quad
M_W = 80.385(15)~\gv,
\quad
M_t = 173.5(1.0)~\gv,
\nonumber \\ &&
G_\mu =1.16637(1)\times 10^{-5}~\gv^{-2}
\;,\;\;\hat{G}_\mu =G_\mu(M_Z)=1.15564(55)\times 10^{-5}~\gv^{-2}
\nonumber \\ &&
\alpha^{-1} = 137.035999\,,\;\;\alpha^{-1}(M_Z^2) = 127.944\,,\;\;
\alpha_s(M_Z^2) = 0.1184(7)\;.
\label{params}
\end{eqnarray}
}
For the Higgs mass we adopt
\begin{eqnarray}
M_H = 125.9\pm 0.4~\gv,
\label{mhinp}
\end{eqnarray}
in accord with latest ATLAS and CMS reports. All light-fermion masses
$M_f\,(f\neq t)$ give negligible effects and do not play any role in
our consideration. The top quark mass given above is taken to be the
pole mass. It should be reminded that it is not precisely clear
whether the value reported by experiments or by the PDG can be
identified with the on-shell mass within the given accuracy. For a
recent review on the subtleties in defining/measuring the top quark
mass see e.g. Ref.~\cite{Juste:2013dsa} and
references therein. The evaluated \MSb parameters may be found in
Table~\ref{tab:params} below.

\section{The SM RG evolution to the Planck scale}
The SM RG in the symmetric phase to two loops has been known for a
long
time~\cite{Jones:1974pg,Jones:1981we,Machacek:1983tz,Machacek:1983fi,Machacek:1984zw}.
More recently important extensions to three loops have been presented
in
Refs.~\cite{Mihaila:2012fm,Bednyakov:2012rb,Bednyakov:2012en,Bednyakov:2013eba,Chetyrkin:2012rz,Chetyrkin:2013wya,Bednyakov:2013cpa}.
Of special interest is the behavior of the Higgs self-coupling
$\lambda$, which plays a key role for the possible stability or
instability of the SM ground state. In fact solutions depend crucially
on including all couplings contributing. For example, it makes a big
difference whether one works in the so called \textit{gaugeless
limit}\footnote{ This term is often used for the approximation
$g'=g=0$. The QCD coupling $g_3$ in any case has to be taken into
account, besides the top Yukawa coupling $y_t$ and the Higgs
self-coupling $\lambda$.}  in the evolution of $y_t$ and $\lambda$ as
in Ref.~\cite{Chetyrkin:2012rz}, for example, or is including also
the gauge coupling contributions as far as they are known (see
e.g. Refs.~\cite{Masina:2012tz,Buttazzo:2013uya} for a fairly complete set of known
corrections). Some time ago RG equations to two loops for the SM
masses as well as for the Higgs VEV in the broken phase have been
calculated in Ref.~\cite{Jegerlehner:2001fb,Jegerlehner:2002er,Jegerlehner:2002em}, where it has been shown that the RG
equations of the symmetric phase are correctly obtained from the ones
in the broken phase. The inclusion of the tadpoles thereby is crucial.

The RG equation for $v^2(\mu^2)$ follows from the RG equations
for the masses and the dimensionless coupling constants using one of the relations
\begin{eqnarray}
 v^2(\mu^2)=4\,\frac{m_W^2(\mu^2)}{g^2(\mu^2)}
=4\,\frac{m_Z^2(\mu^2)-m_W^2(\mu^2)}{g'^2(\mu^2)}
=2\,\frac{m_f^2(\mu^2)}{y^2_f(\mu^2)}
=3\,\frac{m_H^2(\mu^2)}{\lambda(\mu^2)}\epo
\label{vevsquare}
\end{eqnarray}
As a key relation we will use Eq.~(10) of Ref.~\cite{Jegerlehner:2002er}
\begin{eqnarray}
\mu^2 \frac{d}{d \mu^2} v^2(\mu^2)
=3\, \mu^2 \frac{d}{d \mu^2} \left[\frac{m_H^2(\mu^2)}{\lambda(\mu^2)} \right]
\equiv
v^2(\mu^2) \left[\gamma_{m^2}  - \frac{\beta_\lambda}{\lambda} \right]\epo
\label{vev}
\end{eqnarray}
We remind that all dimensionless couplings satisfy the
same RG equations in the broken and in the unbroken phase.

As we know, the Higgs VEV $v$ is a key parameter of the SM, which
interrelates masses and couplings in a well defined way. As a
consequence the RG for mass parameters can be obtained not only by
direct calculation in the broken phase, but also from the knowledge of
the RG of the parameters in the symmetric phase together with the one
for $v(\mu^2)$ or $v^2(\mu^2)=1/\left(\sqrt{2}\,G^{\MSbm}_F(\mu^2)\right)$
as given in Eq.~(\ref{vev}). The proper \MSb definition of a running
fermion mass is
\begin{equation}
m_f(\mu^2) =  \frac{1}{\sqrt{2}}\,v(\mu^2)\,y_f(\mu^2) \;.
\label{eff}
\end{equation}
Of particular interest in our context is the top quark mass for which
the RG equation reads
\begin{eqnarray}
\mu^2 \frac{d}{d \mu^2} \ln m_t^2
= \gamma_t(\alpha_s,\alpha)\epo
\end{eqnarray}
We split $\gamma_t(\alpha_s,\alpha)$ into two parts
$\gamma_t(\alpha_s,\alpha)= \gamma_t^{QCD} + \gamma^{EW}_t \,$, where
$\gamma_t^{QCD}$ is the QCD anomalous dimension, and $\gamma^{EW}_t$
the corresponding electroweak one. $\gamma_t^{QCD}$ includes all terms
which are proportional to powers of $\alpha_s$ only and
$\gamma^{EW}_t$ includes all other terms proportional to at least one
power of $\alpha$, and beyond one-loop multiplied by further powers of
$\alpha$ and/or $\alpha_s$.

In
Refs.~\cite{Jegerlehner:2001fb,Jegerlehner:2002er,Jegerlehner:2002em}
the electroweak contribution to the fermion mass anomalous dimension
$\gamma^{EW}_f$ has been calculated in terms of the RG functions of
the parameters in the unbroken phase of the SM: the result is given by
\begin{eqnarray}
\gamma^{EW}_t & = & \gamma_{y_t} + \frac{1}{2} \gamma_{m^2} - \frac{1}{2} \frac{\beta_\lambda}{\lambda} \;,
\label{SM<->F}
\end{eqnarray}
where  $\gamma_{m^2} \equiv \mu^2 \frac{d}{d \mu^2} \ln m^2 \; ,$
$\beta_\lambda \equiv \mu^2 \frac{d}{d \mu^2} \lambda \;,$
and $ \gamma_{y_q} \equiv \mu^2 \frac{d}{d \mu^2} \ln y_q \;,$
with $y_q$ the quark Yukawa coupling.

In the following we present the results for the running SM parameters
in various plots. The RG equations for the gauge couplings
$g_3=(4\pi\alpha_s)^{1/2}$, $g_2=g$ and $g_1=g'$, for the Yukawa
coupling $y_t$ and for the Higgs potential parameters $\lambda$ and
$\ln m^2$ have been solved in the \MSb scheme with initial values
obtained by evaluating the matching conditions between pole and
running masses. For the case of the dimensionless couplings we
reproduce known results within uncertainties. The \MSb Higgs VEV
square is then obtained by
$v^2(\mu^2)=-\frac{6m^2(\mu^2)}{\lambda(\mu^2)}$ and the other masses
by the relations (\ref{vevsquare}).

\begin{figure}[t]
\centering
\includegraphics[height=5.7cm]{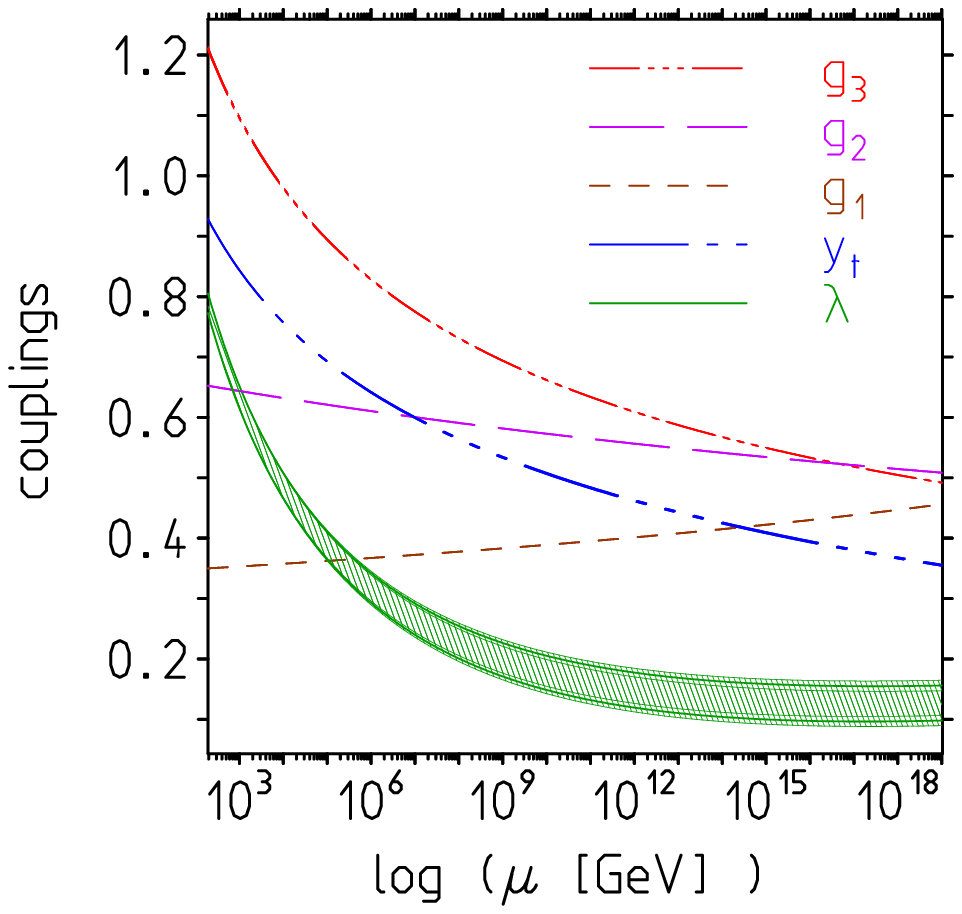}
\includegraphics[height=5.7cm]{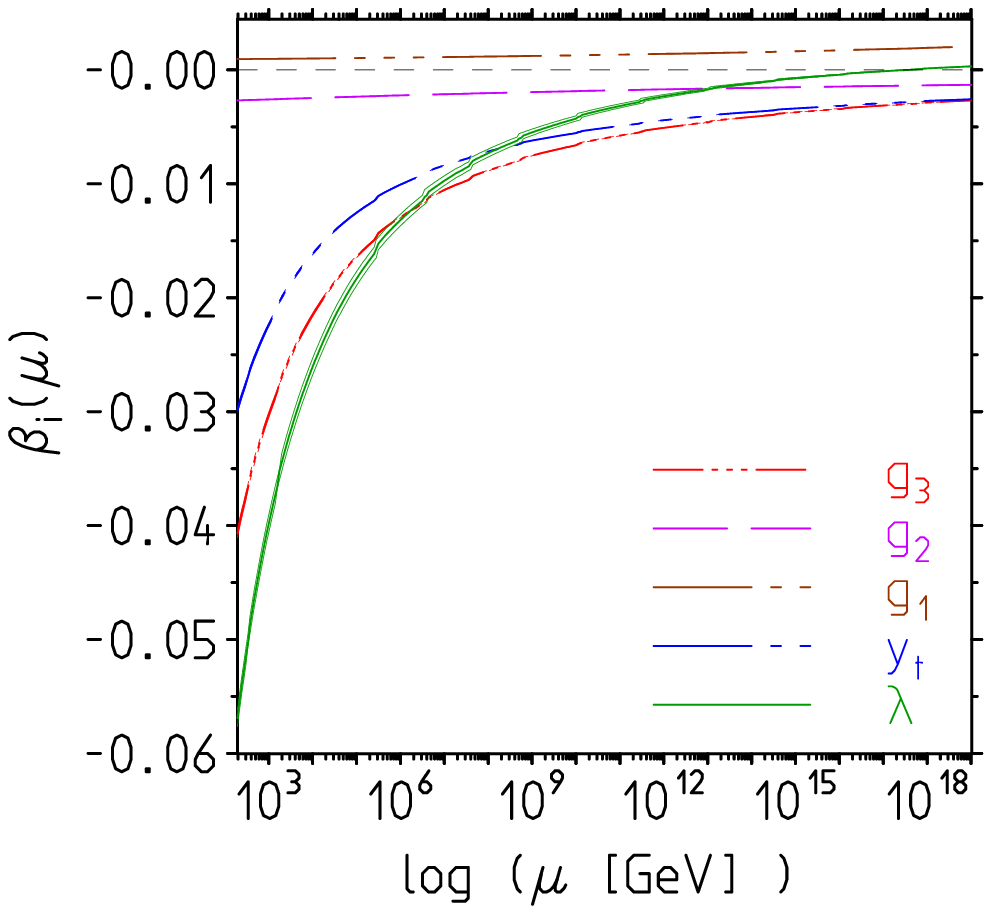}
\caption{Left: the SM dimensionless couplings in the \MSb scheme as a
function of the renormalization scale (see
Refs.~\cite{Yukawa:3,degrassi,Masina:2012tz,Buttazzo:2013uya,Hamada:2012bp}). The input
parameter uncertainties as given in Eqs.~(\ref{params}) and
(\ref{mhinp}) are exhibited by the line thickness. The green shaded band
corresponds to Higgs masses in the range [124-127]~GeV. Right: the
$\beta$-functions for the couplings $g_3$, $g_2$, $g_1$, $y_t$ and $\lambda$.
The uncertainties are represented by the line widths.}
\label{fig:SMrunpar}
\end{figure}

\begin{figure}[t]
\centering
\includegraphics[height=5.7cm]{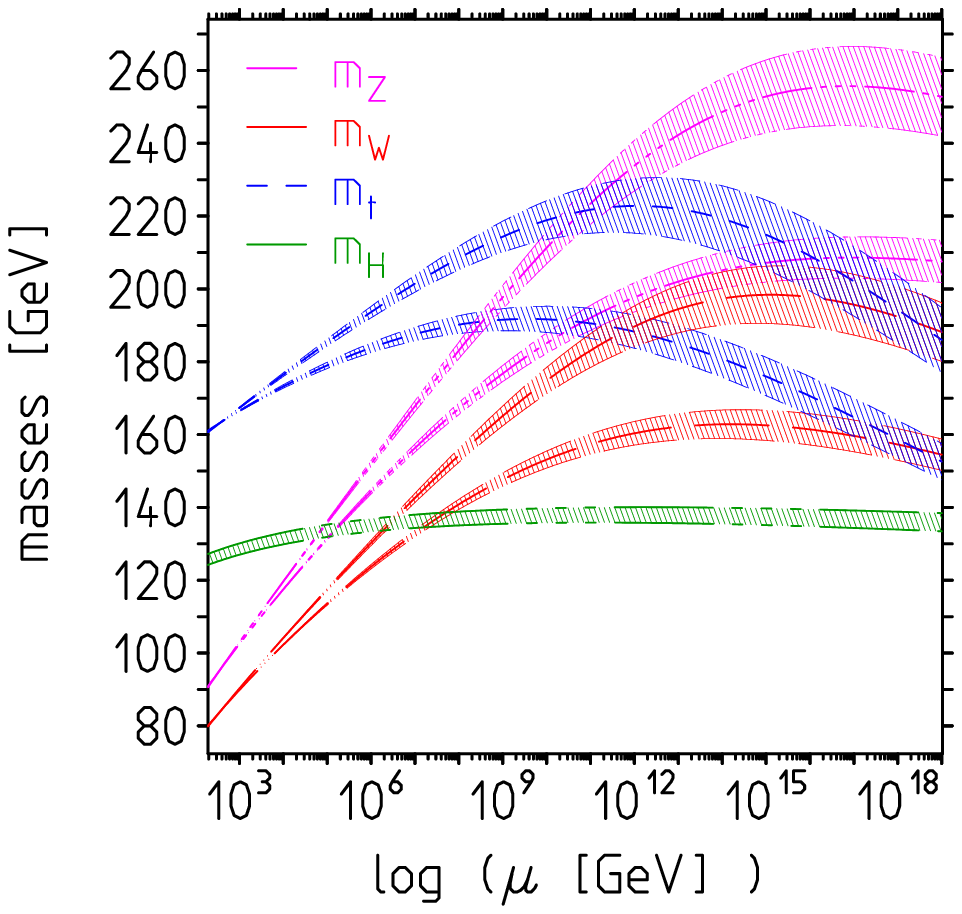}
\includegraphics[height=5.7cm]{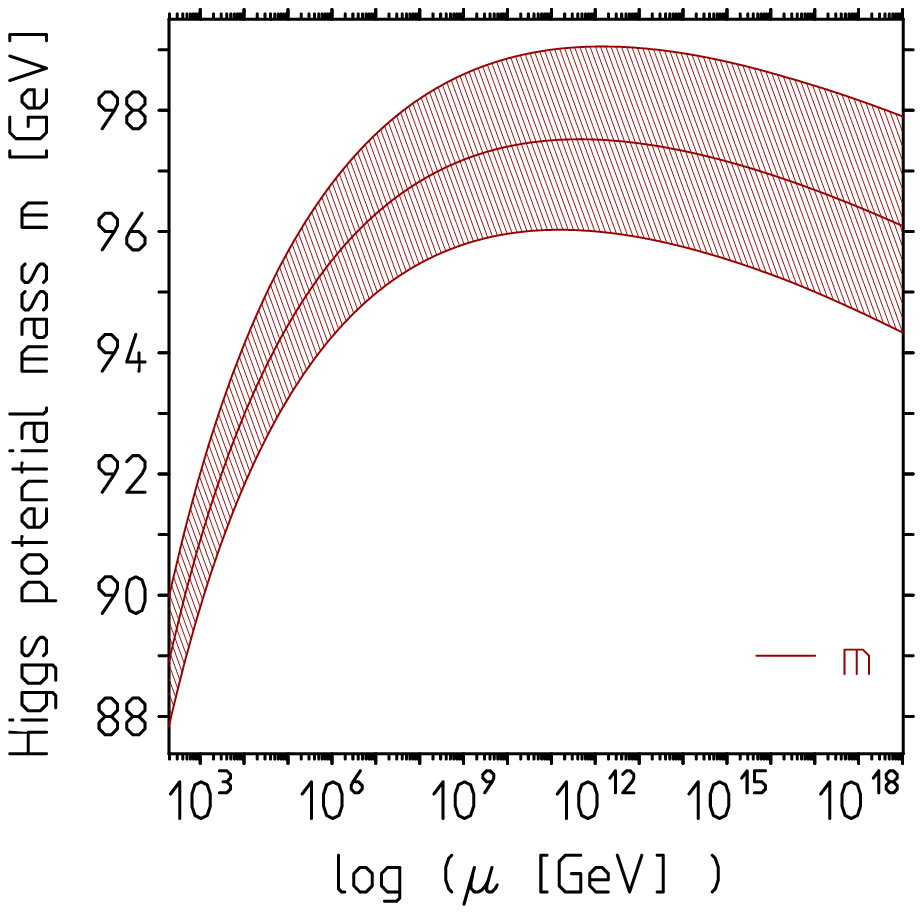}
\includegraphics[height=5.7cm]{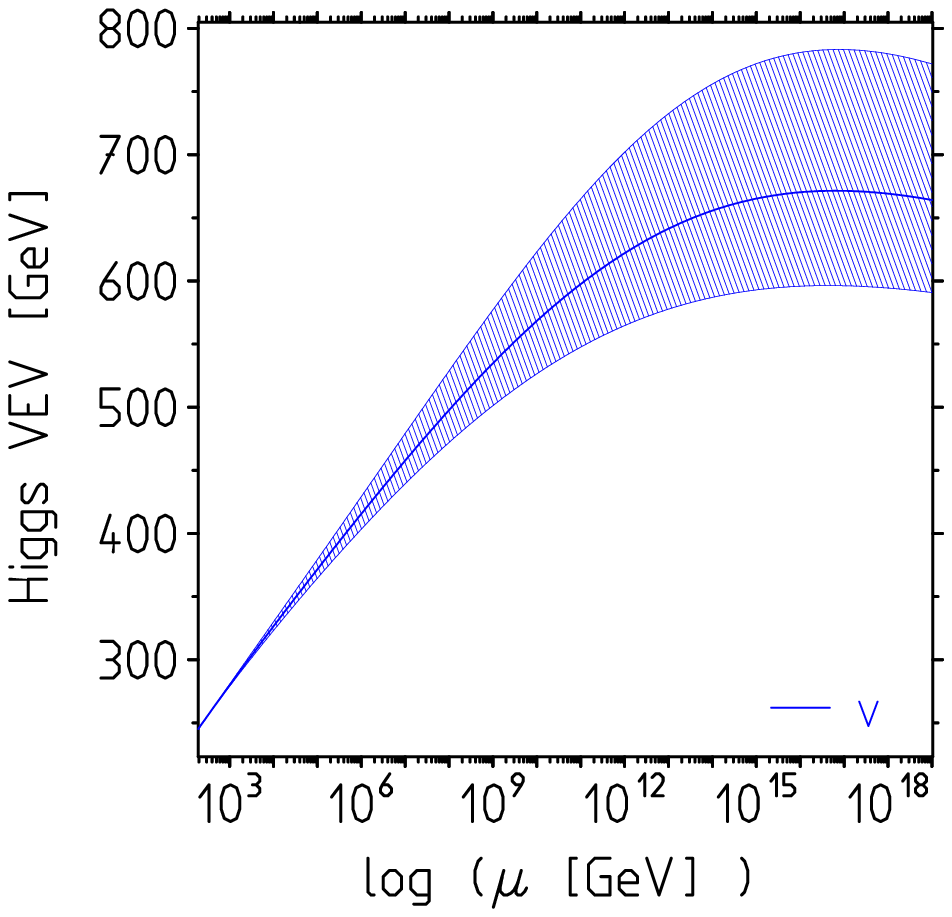}
\includegraphics[height=5.7cm]{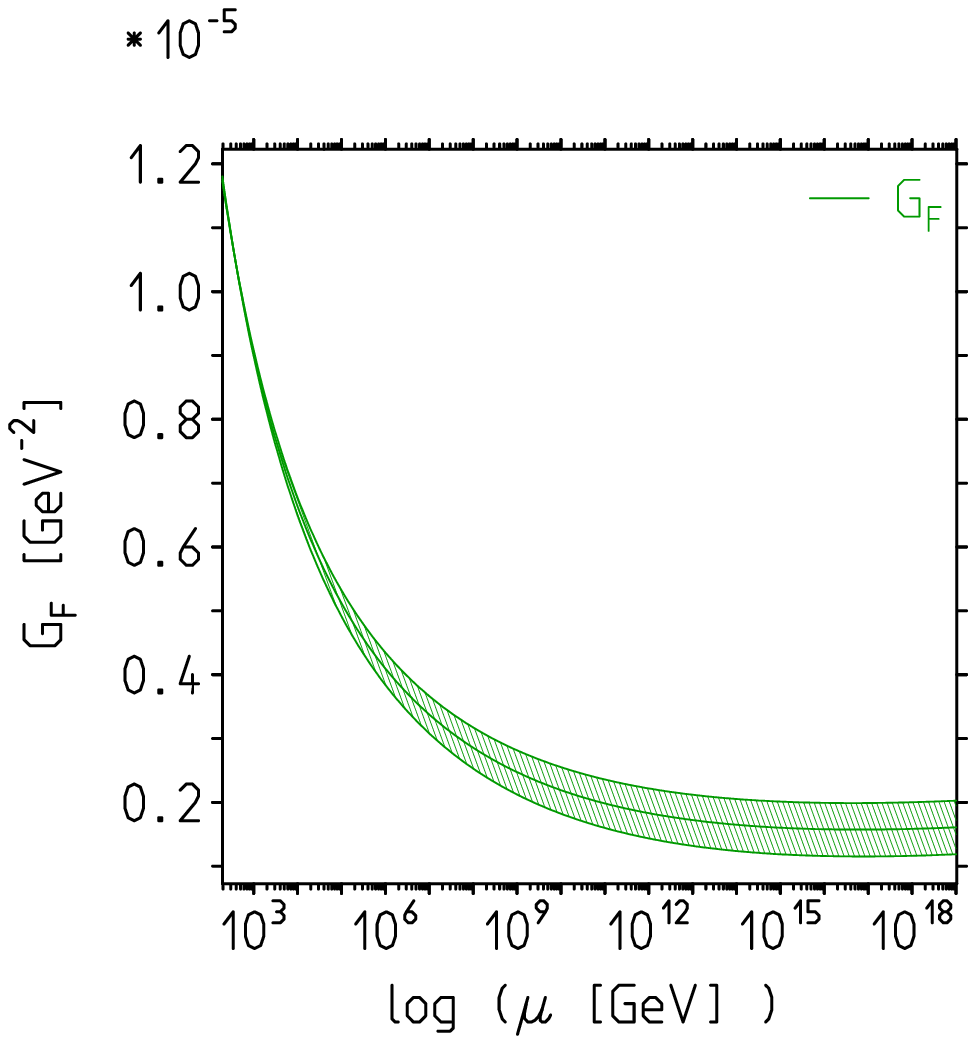}
\caption{Non-zero dimensional \MSb running parameters. Top left: the
running \MSb masses. The shadowed regions show parameter
uncertainties, mainly due to the uncertainty in $\alpha_s$, for a
Higgs mass of 124~GeV, higher bands, and for 127~GeV, lower bands. The
range also determines the green band for the Higgs mass evolution. Top
right: the \MSb Higgs potential parameter $m$. Bottom: $v=\sqrt{6/\lambda}\,m$ and
$G_F=1/(\sqrt2\,v^2)$.  Error bands include SM parameter
uncertainties and a Higgs mass range $125.5\pm1.5~\gv$ which
essentially determines the widths of the bands.}
\label{fig:SMmvpar}
\end{figure}

Figure~\ref{fig:SMrunpar} shows the
solutions of the RG equations and the $\beta$-functions up to
$\mu=\mpl$. The running masses and the
solutions for the Higgs potential mass parameter $m$ as well as $v$ and
the equivalent $G_F$ are
depicted in Fig.~\ref{fig:SMmvpar}

Remarkably, as previously found for the running couplings in
Refs.~\cite{Yukawa:3,degrassi,Masina:2012tz,Hamada:2012bp}, all
parameters stay in bounded ranges up to the Planck scale if one adopts
our matching conditions together with the so far calculated RG
coefficients. With the input parameters evaluated in the previous
section we note that including all known terms no transition to
a metastable state in the effective Higgs potential is observed,
i.e. no change of sign in $\lambda$ occurs. This is in contrast to a
number of other evaluations (which however are not independent as they
essentially use the same input parameters). The difference concerns
the \MSb input-value for the top-quark Yukawa coupling, which in our
case is bases on the analysis Ref.~\cite{Jegerlehner:2012kn}, and
has been confirmed more recently in Ref.~\cite{Bednyakov:2013cpa}.

We observe that the various couplings evolve to values of similar
magnitude at the Planck scale, within a factor of about 2 if we
compare $\sqrt{\lambda}$ with the others. While the gauge couplings
are much closer than they are at low energies, there is no reason for
perfect unification. The different types, the gauge boson-, the
fermion- and the Higgs-couplings have no reason not to differ even if
they emerge form one cutoff system. That the leading couplings are of
the same order of magnitude, however, makes sense in such a kind of
scenario. The emergence of the fermion mass hierarchy is a different
issue.

The key point concerning the behavior of the effective parameters we
may understand when we look at the leading terms of the
$\beta$-functions. At the $Z$ boson mass scale the couplings are given
by $g_1\simeq 0.350$, $g_2\simeq 0.653$, $g_3\simeq1.220$, $y_t\simeq
0.935$ and $\lambda\simeq0.807$. While the gauge couplings behave as
expected, $g_1$ is infrared (IR) free, $g_2$ and $g_3$ are
asymptotically (ultraviolet) free (AF), with leading coefficients
exhibiting the related coupling only, {\small 
$$\beta_1=\frac{41}{6}\,g_1^3\,c\simeq 0.00185\semis
\beta_2=-\frac{19}{6}\,g_2^2\,c\simeq -0.00558\semis
\beta_3=-7\,g_3^3\,c\simeq-0.08049\,, $$}
with $c=\frac{1}{16\,\pi^2}$, the leading top Yukawa $\beta$-function given by
{\small
\begin{eqnarray*}
\beta_{y_t}&=&(\frac92\,y_t^3-\frac{17}{12}\,g_1^2\,y_t-\frac94\,g_2^2\,y_t-8\,g_3^2\,y_t)\,c
\crn &\simeq&  0.02327-0.00103-0.00568-0.07048\crn &\simeq&-0.05391
\end{eqnarray*}
}
not only depends on $y_t$, but also on mixed terms with the gauge
couplings which have a negative sign. In fact the QCD correction is
the leading contribution and determines the behavior. Notice the
critical balance between the dominant strong and the top Yukawa
couplings: QCD dominance requires $g_3>\frac{3}{4}\,y_t$ in the
gaugeless limit.


\noindent Similarly, the $\beta$-function
of the Higgs self-coupling, given by
{\small
\begin{eqnarray*}
\beta_\lambda&=&(4\,\lambda^2-3\,g_1^2\,\lambda-9\,\lambda\,g_2^2+12\,y_t^2\,\lambda+\frac94\,g_1^4
+\frac92\,g_1^2\,g_2^2+\frac{27}{4}\,g_2^4-36\,y_t^4)\,c \crn &\simeq&
        0.01650-0.00187-0.01961+0.05358+0.00021+0.00149+0.00777\crn &&
        -0.17401 \simeq  -0.11595
\end{eqnarray*}
}
is dominated by the top Yukawa contribution and not by the $\lambda$
coupling itself. At leading order it is not subject to QCD
corrections. Here, the $y_t$ dominance condition reads $\lambda<
\frac{3\,(\sqrt{5}-1)}{2}\,y_t^2$ in the gaugeless limit. The top Yukawa
coupling is turned from an intrinsically IR free to an AF coupling by
the QCD term and similarly the Higgs self-coupling is transmuted from
IR free to AF by the dominating top Yukawa term. Including known
higher order terms, except from $\beta_\lambda$, which exhibits a zero
at about $\mu_\lambda\sim 10^{17}~\gv$, all other $\beta$-functions do
not exhibit a zero in the range from $\mu=M_Z$ to $\mu=\MPl$. So, apart
form the $U(1)_Y$ coupling $g_1$, which increases moderately only, all
other couplings decrease and perturbation theory is in good
condition. Actually, at $\mu=\MPl$ gauge couplings are all close to
$g_i\sim 0.5$, while $y_t\sim 0.35$ and $\sqrt{\lambda}\sim 0.36$.

As shown in Fig.~\ref{fig:SMmvpar}, the masses stay bounded up to the
transition point to the symmetric phase, discussed in the next
section. In the broken phase the effective mass relevant for the high
energy behavior, obtained by rescaling all the momenta of the process
$\{p_i\} \to \{\kappa p_i\}$ $\kappa \to \infty$, up to an overall
factor is $ m(\kappa)/\kappa \to 0$ (see Eq.~(\ref{RGsol})).

What is interesting is that the hierarchy of the effective masses gets
mixed up. While the effective Higgs mass $m_H$ and the related Higgs
potential mass $m$ are weakly scale dependent, the Higgs coupling
$\lambda$ drops pretty fast by a factor about 8, together this is
causing the Higgs VEV $v=\sqrt{3/\lambda}\,m_H$ to increase by a
factor about 3.5. Note that, according to the mass-coupling
relationships (\ref{vevsquare}), what compares to the other couplings
is $\sqrt{\lambda}$ not $\lambda$ itself. Given that $m_H$ is weakly
scale dependent, what determines the mass hierarchy are the relations
(see Eq.~(8) of Ref.~\cite{Jegerlehner:2002er})
{\small
\begin{eqnarray*}
\frac{m_W(\mu^2)}{m_H(\mu^2)}=\sqrt{\frac34 \frac{g^2(\mu^2)}{\lambda(\mu^2)}}\;,
\frac{m_Z(\mu^2)}{m_H(\mu^2)}=\sqrt{\frac34 \frac{g^2(\mu^2)+g'^2(\mu^2)}{\lambda(\mu^2)}}\;,
\frac{m_t(\mu^2)}{m_H(\mu^2)}=\sqrt{\frac32 \frac{y^2_t(\mu^2)}{\lambda(\mu^2)}}\,,
\label{runmass}
\end{eqnarray*}
}
which must hold in the broken phase.  Since $g$ is decreasing while
$g'$ is increasing the $Z$ boson mass grows most and exceeds $m_H$
above about $8\power{4}~\gv$ and even $m_t$ above about
$7\power{10}~\gv$. The $W$ boson mass exceeds $m_H$ above about
$5\power{6}~\gv$. These crossings happen in the history of the
universe some time after inflation, Higgs mechanism and EW phase
transition, but long before processes like nucleosynthesis set in. The
effective mass hierarchy is expected to play a role during the EW
phase transition and in some temperature range just below it.

Table \ref{tab:params} lists \MSb couplings at various scales,
representing the central values for $M_H=126~\gv$, which we will use
in the following. Other quark Yukawa couplings are given by 
$y_s(M_t[\mpl])=1.087[0.357]\power{-3}$,
$y_d(M_t[\mpl])=5.151[1.689]\power{-5}$.

\begin{table}[h]
\caption{\MSb parameters at various scales for $M_H=126~\gv$ and
$\mu_0\simeq 1.4\power{16}~\gv$. $C_1$ and $C_2$ are the one- and
two-loop coefficients of the quadratic divergence Eqs.~(\ref{coefC1}) and (\ref{coefC2}),
respectively. Last two columns show corresponding results from
Ref.~\cite{Buttazzo:2013uya}.}  {\begin{tabular}{ccccc||cc}
\hline\noalign{\smallskip}
coupling $\backslash$ scale  & $M_Z$ & $M_t$ & $\mu_0$ & $\mpl$ &
$M_t$~\cite{Buttazzo:2013uya} &
$\mpl$~\cite{Buttazzo:2013uya} \\  \hline
$g_3$ &   $1.2200$ & $1.1644$ & $0.5271$ & $0.4886$ & 1.1644 &\hphantom{-} 0.4873 \\
$g_2$ &   $0.6530$ & $0.6496$ & $0.5249$ & $0.5068$ & 0.6483 &\hphantom{-} 0.5057 \\
$g_1$ &   $0.3497$ & $0.3509$ & $0.4333$ & $0.4589$ & 0.3587 &\hphantom{-} 0.4777 \\
$y_t$ &   $0.9347$ & $0.9002$ & $0.3872$ & $0.3510$ & 0.9399 &\hphantom{-} 0.3823 \\
$y_b$ &   $0.0238$ & $0.0227$ & $0.0082$ & $0.0074$ && \\
$y_\tau$ &$0.0104$ & $0.0104$ & $0.0097$ & $0.0094$ && \\
$\sqrt{\lambda}$&$0.8983$ & $0.8586$ & $0.3732$ & $0.3749$ & 0.8733  & $\I\:\,$0.1131 \\
$\lambda       $&$0.8070$ & $0.7373$ & $0.1393$ & $0.1405$ & 0.7626  &
- 0.0128         \\
$C_1$ & $-6.768$ & $-6.110$ & $\hphantom{0}0\hphantom{.0000}$ & $0.2741$ & $$ & $$\\
$C_2$ & $-6.672$ & $-6.217$ & $\hphantom{0}0\hphantom{.0000}$ & $0.2845$ & $$ & $$\\
$m[GeV]$&$89.096$ & $89.889$ & $97.278$ & $96.498$ &97.278&  \\ \hline
\end{tabular}
\label{tab:params}}
\end{table}
For $M_H=126~\gv$ the zero of $C_1$ is at $\mu_0\simeq 1.4\power{16}~\gv$ the one of $C_2$
at $\mu_0\simeq 1.1\power{16}~\gv\,$. For the same Higgs mass the beta-function $\beta_\lambda$
has a zero at $1.3\power{17}~\gv\,$. Since the difference between $C_1$
and $C_2$ is small we will adopt $C_1$ and the
corresponding value for $\mu_0\,$, in what follows.

\section{The issue of quadratic divergences in the SM}
A discussion of the large scale behavior of the SM is incomplete if
we disregard the problem of quadratic divergences and the related
hierarchy problem.  In contrast to the dimensionless running couplings,
all mass renormalizations (except the photon) are affected by
quadratic ($H$, $W$ and $Z$) or linear divergences (fermions), which
are related universally to the renormalization of the Higgs potential
parameter $m^2$ or equivalently to the Higgs
VEV $v$ in the broken phase.  Standard
\MSb mass RG equations usually only take into account the logarithmic
singularities remaining after ``throwing away'', by analytic
continuation and subsequent $\veps$-expansion, quadratic or linear
divergences. Per se, the RG is a leading log, next-to-leading log, and
so forth resummation tool. Note that this is possible in this way only
in the purely perturbative \MSb scheme, while with a more physical
lattice regularization, which applies beyond perturbation theory, the
quadratic divergences cannot be eliminated this way. In other words
the hierarchy problem is a real problem as reanalyzed recently in
Ref.~\cite{Hamada:2012bp}. In terms of masses the leading one-loop
Higgs mass counterterm is given in Eq.~(\ref{quadraic1}), modulo small
lighter fermion contributions~\cite{Veltman:1980mj}(see also
Ref.~\cite{Jegerlehner:1998kt}). The one-loop coefficient function
$C_1$ may be written as
\bea
C_1=2\,\lambda+\frac32\, {g'}^{2}+\frac92\,g^2-12\,y_t^2\,,
\label{coefC1}
\eea
and is uniquely determined by dimensionless couplings. Surprisingly,
taking into account the running of the SM couplings, which are not
affected by quadratic divergences such that standard RG equations
apply, the coefficient of the quadratic divergences of the Higgs mass
counterterm vanishes at about $\mu_0\sim 1.4 \times 10^{16}~\gv$ given
our set of \MSb input parameters at the scale $M_Z$.  As shown
in Ref.~\cite{Hamada:2012bp} the next-order correction
\bea
C_2&=&C_1+ \frac{\ln (2^6/3^3)}{16\pi^2}\, [
18\,y_t^4+y_t^2\,(-\frac{7}{6}\,{g'}^2+\frac{9}{2}\,g^2
             -32\,g_s^2) \nn \\
             &&-\frac{87}{8}\,{g'}^4-\frac{63}{8}\,g^4 -\frac{15}{4}\,g^2{g'}^2
             +\lambda\,(-6\,y_t^2+{g'}^2+3\,g^2)
             -\frac{2}{3}\,\lambda^2]\,,
\label{coefC2}
\eea
calculated first in Ref.~\cite{Alsarhi:1991ji} (see also Ref.~\cite{Jones:2013aua}),
numerically does not change significantly the one-loop result. The
same result applies for the Higgs potential parameter $m^2$ which
corresponds to $m^2\hat{=}\frac12\,M_H^2$. Thus
\bea
m_0^2=m^2+\delta m^2\;;\;\; \delta m^2= \frac{\Lambda^2}{32 \pi^2}\,C\epo
\label{barem2}
\eea
The relevant parameters are entirely given in terms of SM parameters
in the unbroken phase, which is physical at high energies, as well as
at a different scale in the broken low energy phase, where parameters
are directly accessible. It is important to note that the renormalized
$m^2$ in the symmetric phase is not known and not accessible directly
to experiment, which means that it is not known whether there
is a fine tuning problem in the symmetric phase. As we will see below,
if $m^2$ is not much smaller than the very large $\delta m^2$ it would
affect the inflation pattern and thus in principle is constrained by
the observed properties of Cosmic Microwave Background (CMB)
fluctuations~\cite{PlanckResults}.

For scales $\mu <
\mu_0$ we have $\delta m^2$ large negative, which is triggering
spontaneous symmetry breaking by a negative bare mass
$m_0^2=m^2+\delta m^2$, where $m$ denotes a so-far unknown renormalized
mass.  With increasing energy scale, at $\mu=\mu_0$ the sign of
$\delta m^2$ flips and implies a phase transition to the symmetric
phase, which persists up to the Planck scale.  This means that in the
early universe, up to times about $\sim 0.23 \times 10^{-38}
\mbox{ \ to \ } 10^{-42}$ seconds after the big bang, the SM is 
in the unbroken phase. Finite temperature effects, to be discussed
below, generally are accelerating the transition to the symmetric
phase. This transition is relevant for inflation scenarios in the
evolution of the universe.  At $\mu_0$ the Higgs VEV jumps to zero and
SM gauge boson and fermion masses all vanish, at least provided the
scalar self-coupling $\lambda$ continues to be positive.  Note that
the phase transition scale $\mu_0$ is close to the zero
$\mu_\lambda\sim 1.3\power{17}~\gv$ of the $\beta$-function $\beta_\lambda$
where $\beta_\lambda(\mu_\lambda)=0\,.$ While $\lambda$ is decreasing
below $\mu_\lambda$ it starts to increase weakly above that scale.

The important point is that to all orders of perturbation theory as
well as beyond perturbation theory there exists a solution $C=0$, i.e.
the relation (\ref{quadraic1}) is expected to get corrections from
higher-order effects which are shifting the location of the zero but
do not affect its existence. Such relations are relations between the
dimensionless gauge-, Yukawa- and Higgs-couplings and do not depend on
dimensionful low-energy parameters like the Higgs potential mass $m$
(in the symmetric phase) or the Higgs VEV $v$ (in the broken
phase). Of course $m$, $v$ like $\lpl$ can/must show up as overall
factors in dimensionful quantities. Higher order corrections may
depend on ratios of these. 

\begin{figure}[t]
\centering
\includegraphics[height=7cm]{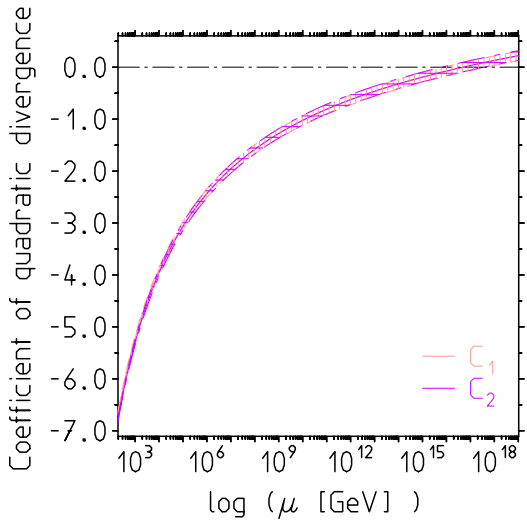}
\caption{The coefficient of the quadratic divergence term at one and
two loops as a function of the renormalization scale. The one-loop
result essentially determines the behavior. The coefficient exhibits a
zero, for $M_H=126~\gv$ at about $\mu_0\sim 1.4 \power{16}~\gv$, not far
below $\mu=\mpl$. The shaded band shows the parameter uncertainties
given in Eqs.~(\ref{params},\ref{mhinp}).}
\label{fig:quaddivcoef}
\end{figure}

\begin{figure}[t]
\centering
\includegraphics[height=5.7cm]{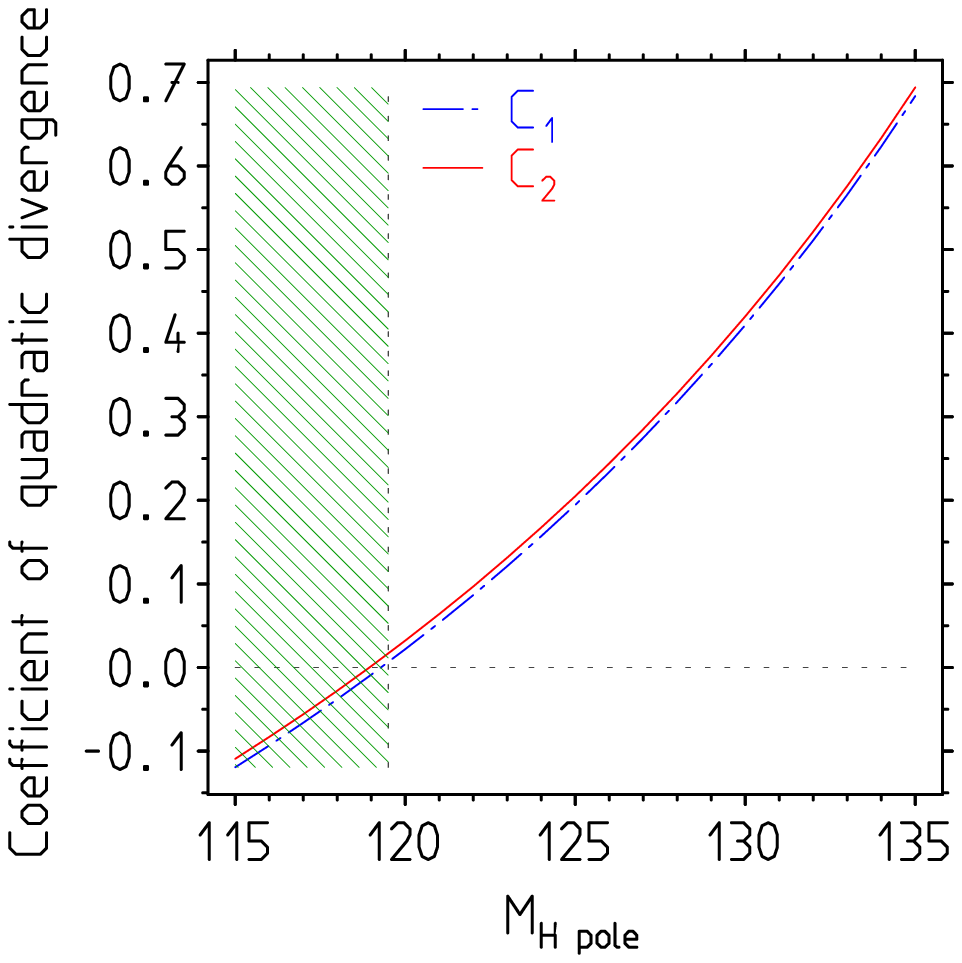}
\includegraphics[height=5.7cm]{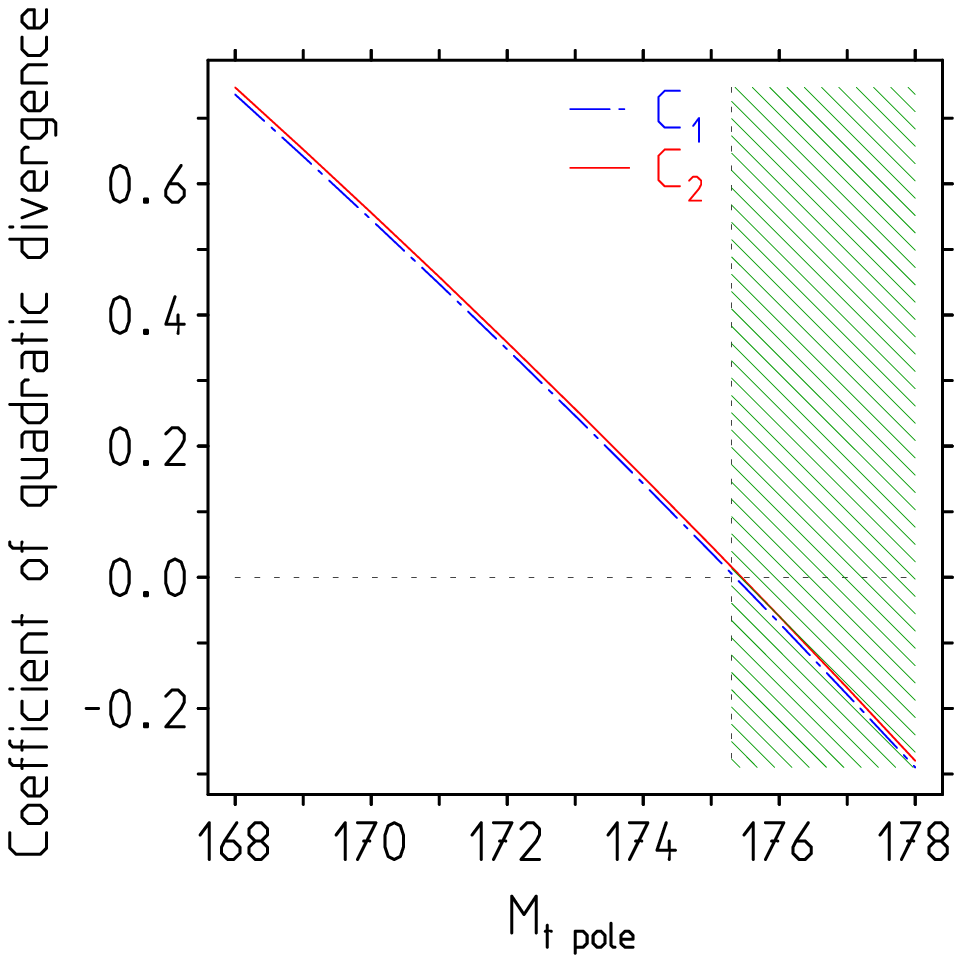}
\caption{Left: the coefficient of the quadratic divergence term
at $\mu=\mpl$ as a function of $M_H$ for $M_t=173.5~\gv$. Right: the same as a
function of $M_t$ for $M_H=125~\gv$. In the shaded region the zero is
above the Planck scale and thus unphysical in our LEESM context.}
\label{fig:quaddivatMPl}
\end{figure}

\begin{figure}[t]
\centering
\includegraphics[height=5.7cm]{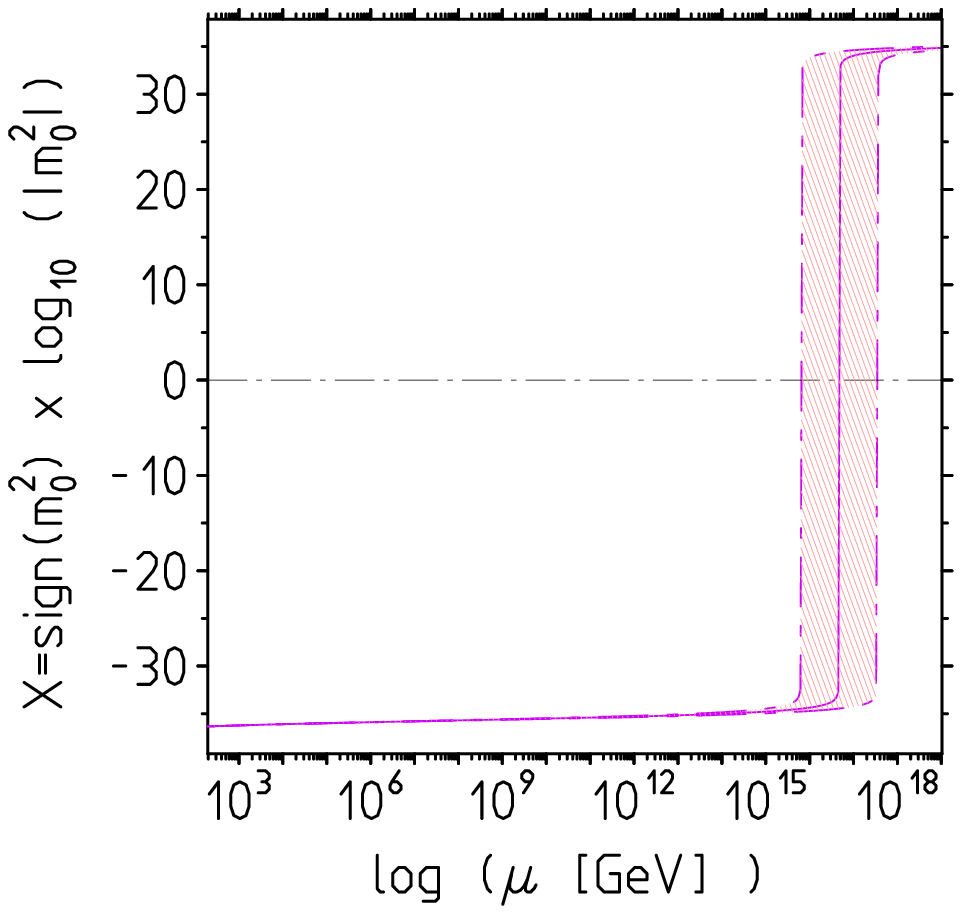}
\includegraphics[height=5.7cm]{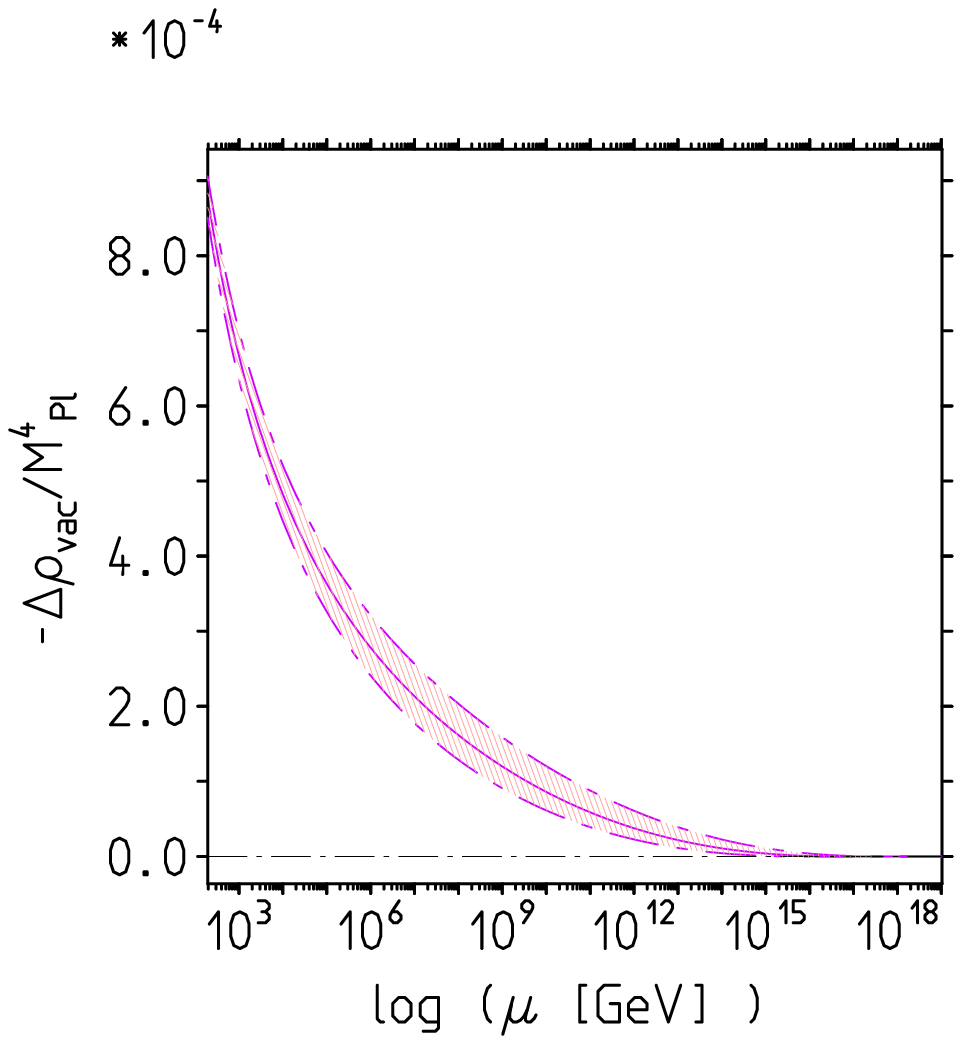}
\caption{The Higgs phase transition in the SM. Left: shown is $X=\sign(
m^2_0)\times \log_{10} (|m^2_0|)$ which represents
$m^2_0=\sign(m^2_0)
\times 10^{X}$. In the broken phase $m^2_0=\frac12\,m^2_{H0}$. At the zero of the coefficient
function shown in Fig.~\ref{fig:quaddivcoef} the counterterm $\delta
m^2=m^2_0-m^2=0$ ($m$ the \MSb mass) vanishes and the bare
mass changes sign. The band represents the parameter uncertainties
Eqs.~(\ref{params},\ref{mhinp}). Right: the ``jump'' $-\Delta
\rho_{\rm vac}=\frac{\lambda}{24}\, v_0^4$ in units of
$\mpl^4$ as a function of the renormalization scale $\mu\,$. The jump,
too small to be seen in this plot, agrees with the renormalized one.}
\label{fig:deltam2}
\end{figure}

Given all masses of the SM we note that $4 M_t^2 > M_H^2+M_Z^2+2
M_W^2$ which makes the Higgs mass counterterm $\delta M_H^2 < 0$ and
it is the heavy top quark which triggers spontaneous symmetry breaking
in the SM as the bare mass square $m_{H0}^2 = M_H^2
+ \delta M_H^2$ is driven to negative
values. Figure~\ref{fig:quaddivcoef} shows how the coefficient $C$ is
dominated by the leading order term and as a function of the \MSb
running couplings vanishes and changes sign below the Planck scale.
Interestingly, the top quark mass and the Higgs mass in conspiracy
with the other relevant couplings are such that the quadratic
divergence vanishes precisely not far below the Planck scale, as
illustrated in a different way in Fig.~\ref{fig:quaddivatMPl}. The
observation that, taking into account the scale dependence, there is a
zero in the coefficient $C_i$ ($i=1,2$) of the quadratic divergence
has been pointed out in Ref.~\cite{Hamada:2012bp}. Also in this case,
the precise location of the zero is sensitive in particular to the 
top-quark Yukawa coupling and with their input, in
Ref.~\cite{Hamada:2012bp}, the zero was found to be located above the
Planck scale, which in our LEESM scenario would not have a physical
meaning.
Concerning the hierarchy problem, a zero at some scale does not
eliminate the problem altogether of course, as illustrated by
Fig.~\ref{fig:deltam2}. Nevertheless, in the vicinity of the
transition point the system seems not to remember the short distance
scale $\lpl$ and the VEV which develops at the close-by EW phase
transition takes a value which need not be related to the short
distance scale. 

Here we have another argument why $v$ can have any value we want: let
us consider the location of the minimum of the potential as a function
of the Higgs potential parameters $m_0^2$ and $\lambda\,.$ For given
positive $\lambda$, when we vary $m^2_0$ from positive to negative
values as it happens when changing the energy scale $\mu$ from above $\mu_0$
to below it. One has $v\equiv0$ when $m_0^2\geq0$ and
$v^2=-\frac{6m_0^2}{\lambda}$ as soon as $m_0^2<0$. So $v$ is a continuous
function of $m_0^2$ and can take any value. It is certainly not
justified to assume that $v$ is jumping from zero to $\mpl$ suddenly.

Another point concerning the Higgs transition\footnote{We use the
term ``Higgs transition point'' for the point where the Higgs mechanism 
would take place in the zero temperature SM. The Higgs transition
point lies above the EW phase transition point because of finite
temperature effects which must be taken into account when considering
the evolution of the hot early universe (see below).} and the meaning of the
key relation (\ref{barem2}). Before the Higgs mechanism has taken
place in the cooling down of the universe we are at very high energy
and we see the bare theory. At these scales a relation like
(\ref{barem2}) is observable, i.e. all three terms have a physical
meaning and in principle are accessible to experiments. Below the
Higgs transition we are in the low energy regime characterized by the
long range quantity $v$, which results form long range collective
behavior of the system. In this case the relation (\ref{barem2})
provides a matching relation between renormalized and bare quantities
in the broken phase, for $\delta m^2=0\,.$ In the low energy phase the
bare $m_0^2$ and the counterterm $\delta m^2$ are not observable any
longer. The relation is not testable by low energy experiments, and if
we try to test it by short distance experiments we are back testing
the symmetric phase where the quantities have changed their values and
meaning.

At the transition point $\mu_0$ we have $v_0=v(\mu_0^2)$, where $v(\mu^2)$ is the \MSb renormalized VEV. Thus
the contribution to the vacuum energy \bea
\Delta \rho_{\rm
vac}=-\frac{\lambda(\mu^2_0)}{24}\, v^4(\mu_0^2)\,,
\eea 
is $O(v^4)$ and {\bf not} $O(\mpl^4)\,.$
In any case, depending on finite temperature effects, near the zero of
the coefficient function there is a phase transition, which
corresponds to a restoration of the symmetry. Taking into account
input parameter uncertainties, the transition is found to take place
at a scale in the range $\mu\sim 10^{16} \mathrm{ \ to \ }
10^{18}~\gv$, also depending on finite temperature effects. This is
one to three orders of magnitude below the Planck scale. Now, in the
symmetric phase, the positive quadratically enhanced bare mass term
has the potential to trigger inflation. Note that at the zero of
$\beta_\lambda$ at about $\mu_\lambda \sim 1.3 \power{17}~\gv > \mu_0$
the Higgs self-coupling $\lambda$ although rather small is still
positive and then starts slowly increasing up to $\mpl$. The point
$\mu=\mu_\lambda$, where $\beta_\lambda(\mu_\lambda) =0$, corresponds
to a phase transition form the antiscreening to the screening
phase. It is not a RG fixed point though, because the
$\beta_\lambda(\mu)$--function depends on other couplings which also
change with the scale.

We also note that a zero of $\lambda$ in the Higgs phase formally lets
the Higgs VEV $v$ explode: $v^2(\mu^2)=-6\,m^2(\mu^2)/\lambda(\mu^2)
\to \infty$ as $\lambda(\mu^2)\to 0$. Several analyses (see
Ref.~\cite{Buttazzo:2013uya} and references therein), which find a
somewhat higher \MSb input value for $y_t(M_t^2)$ find a zero of
$\lambda$ as low as $\mu \sim 10^{9}~\gv$.  Except for the Higgs mass
$m_H=\sqrt{2}\,m$ all masses would reach values $O(\mpl)$. In the
LEESM scenario of course higher dimensional operators would save
stability of the potential, which is assumed to be a given property of
the Planck medium. It would mean that dimension 6 operators come into
play at much lower scales than expected by naive $E/\Lambda_{\rm Pl}$
counting.

If we take renormalization as a physical process, similar to what it is
in condensed matter physics, where both bare and renormalized (effective)
quantities are physical and accessible to experiments, the key
question is what happens to the effective Higgs potential
$V=\frac{m_0^2}{2}\,H^2+\frac{\lambda}{24} H^4$. When the $m_0^2$-term
changes sign and $\lambda$ stays positive, we know it is a first order
phase transition. The latter term maybe is used somewhat sloppy
here. Actually, when continuously lowering the temperature coming
from the high energy side $m_0^2(\mu)$ is a continuous function of
$\mu$.  If we would have $m^2=0$, i.e. $m_0^2=\delta m^2)$, then
$v_0(\mu)$ ($v_0^2=-\frac{6m_0^2}{\lambda}$ when $m_0^2<0$) would be a
continuous function of the scale as well, although a non-analytic
one. When $\mu\geq \mu_0$ $v_0(\mu)\equiv 0$ and for $\mu<\mu_0$ we
have $v_0(\mu)>0$ monotonically increasing as $\mu$ decreases. Thus
the point $\mu_0$ is the end point from below of a continuous family
of first order transition points and hence itself represents a second
order phase transition point. Actually, because the renormalized
$m^2>0$ is non-zero, more precisely, the Higgs transition point
$m_0^2=0$ is reached when $\delta m^2=-m^2$, which is slightly below
$\mu_0$ at $\mu_H$. However, the ``high energy meets low energy''
matching point is $\delta m^2=0$ where $m_0^2=m^2$. Here, the
renormalized \MSb mass square $m^2=m^2(\mu_0)$ is given, fixed via RG
running and low energy \MSb vs. on-shell matching condition by the
experimental value of the Higgs mass $M_H$. Note that the initial
condition $m^2=m^2(\mu_0)$ is a point still in the symmetric phase and
when the Higgs transition takes place at $\mu_H<\mu_0$ we have a
finite $v^2(\mu_H)=-\frac{6m^2(\mu_H)}{\lambda(\mu_H)}$,
i.e. factually we see a jump from $v=0$ to $v\neq 0$ at the Higgs transition. In that
sens the Higgs transition represents a first order phase transition.

Since the bare Lagrangian is the true Lagrangian (renormalization is
just reshuffling terms) the change in sign of the bare mass is what
determines the phase\footnote{In the broken phase we have the mass
coupling relations (\ref{runmass}), which also must hold for the bare
parameters. These relations then tell us that linear (fermion masses)
and quadratic (boson masses) divergences are absent at the transition
point. Except for $m$ all other masses ideally vanish in the symmetric
phase.}.

Above the transition point the number of massless degrees of freedom
(radiation) of the SM consists of $g_f=90$ fermionic degrees of
freedom and $g_B=24$ bosonic ones such that the effective number of
degrees of freedom 
\bea g_*(T)=g_B(T)+\frac78\,g_f(T)=102.75
\eea 
(the factor $\frac78$ accounts for the Pauli exclusion principle which
applies for the fermions). The four Higgses in the symmetric phase
have equal masses, and are very heavy. If all SM modes would be
massless we would have $g_*(T)=106.75$, which effectively applies for
temperatures large relative to about $2\,M_{\rm max} < 500~\gv$, where
$M_{\rm max}$ is the upper bound for all running SM masses in the
range up to the transition point (see Fig.~\ref{fig:SMmvpar}). Below
the transition point we know that the one remaining physical Higgs is
as light as the other SM particles\footnote{Highly relativistic particles then
contribute $\rho_{\rm rad}(T)=\frac{\pi^2}{30}\,g_*(T)\,T^4$ to the
radiation density.}. This shows that it need not be true that the
higher the energy the more relativistic degrees of freedom must show
up. The reason of course is that we crossed  phase transition line.

Since the Higgs phase transition and inflation happen very early in
the thermal history of the universe at times when the universe is very
hot and dense (hot big bang), finite temperature effects must be
included in a realistic treatment of the EW phase transition and
inflation~\cite{FiniteTemp1,FiniteTemp2,FiniteTemp3,Dine:1992wr}. The
leading modification caused by finite temperature effects enters the
finite temperature effective potential $V(\phi,T)$: while at zero
temperature $V(\phi,T=0) = -\frac{\mu^2}{2}\,\phi^2+\frac{\lambda}{24}
\,\phi^4\,,$ at finite temperature we have 
\bea
V(\phi,T) =
\frac{1}{2}\,\left(g_T\,T^2-\mu^2\right)\,\phi^2+\frac{\lambda}{24}
\,\phi^4+ \cdots \,. 
\eea
Usually it is assumed that the Higgs is in the broken phase
($\mu^2>0$) and that the EW phase transition is taking place when the
universe is cooling down below the critical temperature
$T_c=\sqrt{\mu^2/g_T}$. However, above the scale $\mu_0$ we are in the
symmetric phase with $-\mu^2\to m_0^2=m^2+\delta m^2>0\,.$ As
claimed before, the phase transition is triggered by $\delta m^2$
with $m_0^2\simeq1.74\power{-3}\,\mpl^2\,.$ In our case we have
$T(\mu=\mu_0)\simeq 1.62\power{29}~\degK$ and $T(\mu=\mpl)\simeq
1.42\power{32}~\degK$ such that we expect the EW phase transition to
be triggered by the bare Higgs mass in spite of the fact that the
finite temperature term $g_T\,T^2$ is very large in the early
universe. The SM coefficient $g_T$ is given by~\cite{Dine:1992wr}
\bea
g_T=\frac{1}{4v^2}\,\left(2m_W^2+m_Z^2+2m_t^2+\frac12\,m_H^2\right)
=\frac{1}{16}\,\left[3\,g^2+{g'}^{2}+4\,y_t^2+\frac23\,\lambda
\right]\,,
\eea
and we can calculate its value near $\mpl$ given the effective couplings at $\mpl$
listed in Table~\ref{tab:params}. We estimate $g_T \approx 0.0980$. Therefore, near above the phase transition point, where
$m^2_0=m^2(\mu^2_0)$, the bare mass term is
dominating. Up at the Planck scale the temperature term is expected to
dominate in general. 
\begin{figure}[t]
\centering
\includegraphics[height=5.8cm]{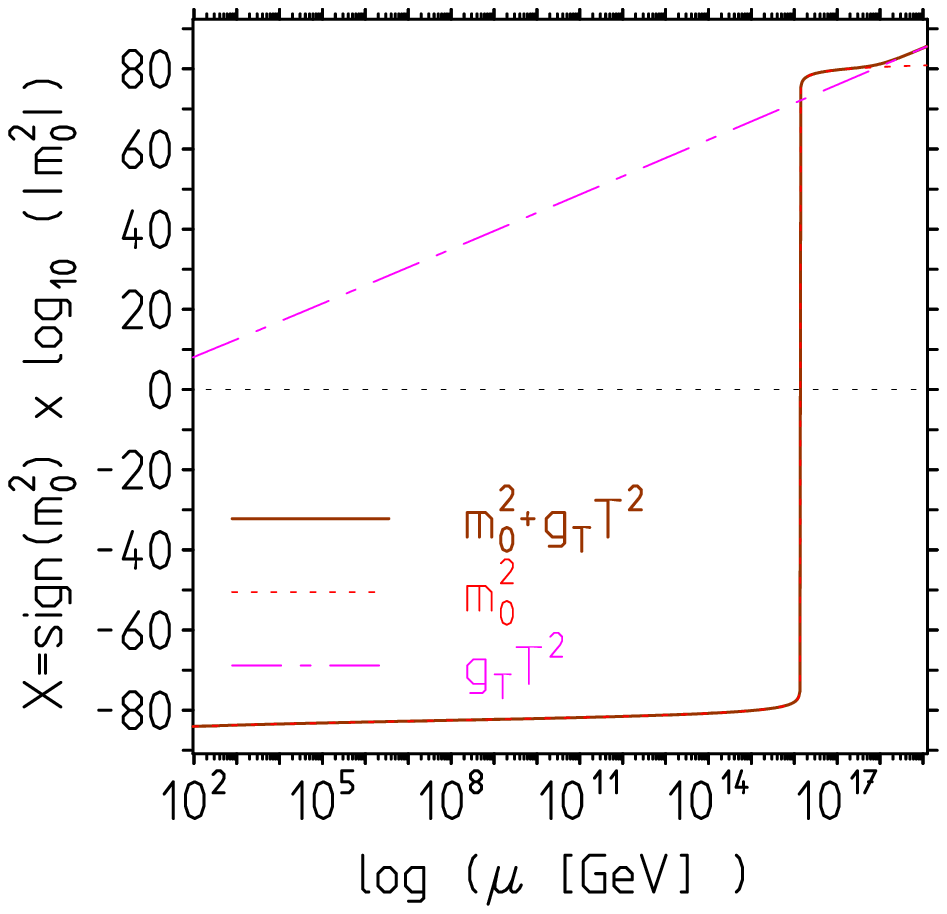}
\includegraphics[height=5.8cm]{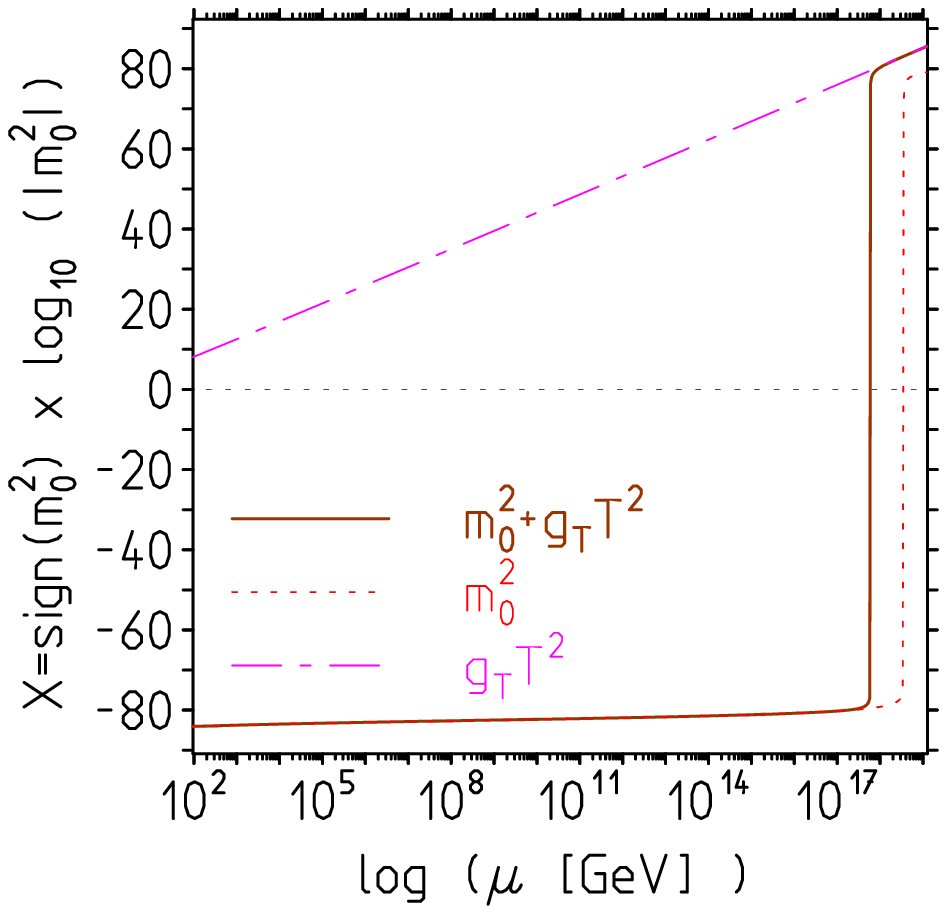}
\caption{The role of the effective bare mass in the finite temperature
SM. Left: for $\mu_0\sim 1.4 \power{16}~\gv$ ($M_H\sim
126~\gv$, $M_t\sim 173.5~\gv$).
Right: finite temperature delayed transition for
$\mu_0\sim 6 \power{17}~\gv$ ($M_H\sim 124~\gv$, $M_t\sim 175~\gv$),
the $m^2_0$ term  alone is flipping at about $\mu_0\sim 3.5 \power{18}~\gv$.}
\label{fig:FT}
\end{figure}
However, this depends on the value of the
renormalized $m^2$-term in the symmetric phase. Here and in the following we \textbf{assume}
that the bare mass is dominated by the quadratically enhanced mass
counterterm, meaning we consider the renormalized $m^2$ to satisfy
$m^2\ll \delta m^2$. Inflation as well
as EW phase transition scenarios certainly depend on this
assumption. As $\delta m^2(\mu)$ is a running mass, which
vanishes at $\mu_0$, the finite renormalized $m^2$ could come into
play at a late stage of inflation, and could be related to the value
$m\sim 10^{-6}\,\mpl$ which has been extracted form observed inflation
properties as a preferred scalar mass in standard inflation
scenarios. Such a finite addition would not affect much the Higgs
transition and the subsequent EW phase transition. In any case, the
phase transition seems to be triggered quite generally by the sign
flip of the bare mass term, as is illustrated in
Fig.~\ref{fig:FT}. The EW phase transition can take place only after
the Higgs mass flip. Of course, our rough estimates are no substitute
for a more careful reanalysis of the EW phase transition.

Let us finally consider the behavior of $v(\mu)$ in some more detail.
The crucial point is that the running of $v(\mu)$ is determined by the
anomalous dimension of the Higgs potential parameter $m^2$ and by the
$\beta$-function related to the renormalization of $\lambda$. Its
behavior has been investigated recently in Ref.~\cite{Jegerlehner:2012kn}. For high
energies the second term of (\ref{vev}) is dominating, such that
\begin{equation*}
\mu^2 \frac{d}{d \mu^2} \ln v^2(\mu^2)  \sim - \frac{\beta_\lambda(\mu)}{\lambda(\mu)} \;.
\end{equation*}
The behavior of $\lambda(\mu)$ and $\beta_\lambda(\mu)$ has been
studied recently in the context of vacuum stability of the SM Higgs
sector in
Refs.~\cite{Yukawa:3,degrassi,Moch12,Mihaila:2012fm,Chetyrkin:2012rz,Masina:2012tz,Chetyrkin:2013wya}
and reveals that the beta function $\beta_\lambda$ is negative up to a
scale of about $1.3 \times 10^{17}~\gv$, where it changes sign. As already
mentioned, above the zero $\mu_\lambda$ of $\beta_\lambda$, the
effective coupling starts to increase again and the key question is
whether at the zero of $\beta_\lambda$ the effective coupling is
still positive. In the latter case it will remain positive although
small up to the Planck scale. In any case, at moderately high scales
where $\beta_\lambda <0$, and provided $\lambda$ is still positive the
following behavior is valid for the Higgs VEV:
\begin{equation}
\left. v^2(\mu^2) \right|_{\mu^2 \to \infty}
\sim  \left(\mu^2\right)^{-\frac{\beta_\lambda(\mu)}{\lambda(\mu)}}
\to \infty \;,
\label{RG-PLANK}
\end{equation}
which means that $v^2(\mu^2)$ is increasing at these
scales (where $\beta_\lambda <0$ and $\lambda>0$). The
analyses Refs.~\cite{Yukawa:3,degrassi,Moch12,Mihaila:2012fm,Chetyrkin:2012rz,Masina:2012tz,Chetyrkin:2013wya,Buttazzo:2013uya} find that $\lambda$ turns negative (unstable
or meta-stable Higgs potential) before the beta function reaches its
zero. This may happen at rather low scales around $10^{9}~\gv$. Some
consequences we have mentioned above. In our approach a zero of $\lambda$ below the Planck
scale would represent an essential singularity. 
According to our analysis $\lambda$ remains positive up to the zero of
the beta function and as a consequence up to the Planck scale in
agreement with Refs.~\cite{Jegerlehner:2012kn,Bednyakov:2013cpa}.

If $\lambda=0$ before $\beta_\lambda=0$ the SM cannot be valid
beyond that point $\mu^*$ where $\lambda(\mu^*)=0$. Note that in a
renormalizable theory renormalization does not induce
non-renormalizable higher-order terms, so one really has to give up
the SM in its literal form. Here one has to remember that the SM is an
effective theory only and at higher scales non-renormalizable
operators must come into play. The next relevant term would be
$\frac{\xi}{\Lambda_{\rm Pl}^2}\frac{1}{6!}H^6$ with
positive dimensionless coupling $\xi$, which keeps the system stable. Then one has
three parameters in the relevant part of the potential $m^2$,
$\lambda$ and $\xi$ and one may have more complicated vacuum
structure, with metastable states etc. In other words the case
$\lambda<0$ cannot be discussed without extending the SM.


\section{The impact on inflation}
As the Higgs system persists to make sense back to times of the early
universe, it is attractive to think that the SM Higgs field itself is
responsible for the inflation era of the early universe, as originally
thought by Guth~\cite{Guth:1980zm} (see also
Refs.~\cite{Starobinsky:1980te,Linde:1981mu,Albrecht:1982wi,Mukhanov:1981xt,Mukhanov:1985rz,Mukhanov:1990me}).
Major phenomenological input on inflation comes form Cosmic Microwave
Background observation, most recently from the Planck mission
(see Ref.~\cite{PlanckResults} and references therein). The
``inflation term'', which comes in via the SM energy-momentum tensor,
adds to the r.h.s. of the Friedmann equation
\begin{eqnarray}
\ell^2\,\left(V(\phi)+\frac12 \,\dot{\phi}^2\right)\;,
\end{eqnarray}
where $\ell^2=8\pi G/3$. $M_\mathrm{\rm Pl}=(G)^{-1/2}$ is the Planck
mass, $G$ is Newton's gravitational constant and for any quantity $X$
we denote time derivatives by $\dot{X}$. In this section dealing with
physics near the Planck scale $\phi,V(\phi),\lambda$ and $m$ denote
the bare quantities (fields and parameters). Inflation requires an
exponential growth $a(t)\propto
\E^{Ht}$ of the Friedman-Robertson-Walker radius $a(t)$ of the universe, where
$H(t)=\dot{a}/a(t)$ is the Hubble constant at cosmic time $t$.

The contribution of the Higgs to the energy momentum tensor amounts
to a contribution to energy density and pressure given by
\begin{eqnarray}
\rho_\phi=\frac12\,\dot{\phi}^2+V(\phi)\semis
p_\phi=\frac12\,\dot{\phi}^2-V(\phi)\epo
\end{eqnarray}
The second Friedman equation has the form
$\ddot{a}/a=-\frac{\ell^2}{2}\,\left(\rho+3p\right)$ and the
condition for growth $\ddot{a}>0\,,$ requires $p<-\rho/3$ and hence $\frac12
\dot{\phi}^2< V(\phi)$. CMB observations strongly favor the
slow-roll inflation $\frac12\dot{\phi}^2\ll V(\phi)$ condition and
favors the dark energy equation of state $w=p/\rho=-1$. Indeed the Planck mission
measured $w=-1.13^{+0.13}_{-0.10}$. The
first Friedman equation reads $\dot{a}^2/a^2+k/a^2=\ell^2\,\rho$ and
may be written as $H^2=\ell^2\,
\left[V(\phi)+\frac12\,\dot{\phi}^2\right]=\ell^2\,\rho$. The kinetic 
term $\dot{\phi}^2$ is controlled by $\dot{H}=-\frac32
\ell^2\,\dot{\phi}^2$ related to the observationally controlled
deceleration parameter $q(t)=-\ddot{a}a/\dot{a}^2$. In addition we have the field
equation 
\bea
\ddot{\phi}+3H\dot{\phi}=-V'(\phi)\equiv -\D V(\phi)/\D \phi\,.
\label{fieldeq}
\eea
By inflation
$k/a^2(t) \to 0$ ($k=0,\pm 1$ the normalized curvature), such that the
universe looks effectively flat ($k=0$) for any initial $k$. Inflation
looks to be universal for quasi-static fields $\dot{\phi}\sim 0$ and
$V(\phi)$ large positive. Then $a(t)\propto
\exp (Ht)$ with $H\simeq
\ell \sqrt{V(\phi)}$. This is precisely what the SM Higgs system in the symmetric phase
suggests, if in the Higgs potential $\lambda$ remains positive
and the bare mass square $m^2$ is positive too. As both $\lambda$ and
$m^2$ for the first time are numerically fairly well known,
quantitative conclusions on the inflation patterns should be possible
solely on the basis of SM properties. The leading behavior is
characterized by a free massive scalar field with potential
$V=\frac{m^2}{2}\,\phi^2$, such that
$H^2=(\dot{a}/a)^2=\ell^2\,\frac{m^2}{2}\,\phi^2$ and
$\ddot{\phi}+3H\dot{\phi}=-m^2\phi\,,$ which is nothing but a harmonic
oscillator with friction. A constant background field $\phi \to
\phi_0+\phi$ would imply a dark energy term (cosmological constant) of
the right sign. In contrast, after the phase transition triggered by the
change of sign in the bare $m^2$, the scalar VEV implies a cosmological
constant contribution $-\frac{\lambda}{24}\, v^4$ of negative
sign.

Note that, as required by the CMB horizon
problem, the exponent $Ht$ is much larger than unity
if \mbo{\phi} exceeds the Planck mass at these times. Needed is
\mbo{N_e \simeq H t > 60} to solve the horizon
problem. The inflation blow-up exponent is given by
\bea
N_e&=&\ln \frac{a(t_{\rm end})}{a(t_{\rm
initial})}=\int\limits_{t_i}^{t_e}\,H(t)\,\D t
=\int\limits_{\phi_i}^{\phi_e}\,\frac{H}{\dot{\phi}}\,\D \phi
\nn \\
&=&-\frac{8\pi}{\mpl^2}\int\limits_{\phi_i}^{\phi_e}\,\frac{V}{V'}\,\D
\phi =H\,(t_e-t_i)\;,
\label{Nedef}
\eea
and \mbo{N_e=H\,(t_e-t_i)} is exact if \mbo{H=\mathrm{constant}}
i.e. when \mbo{\rho=\rho_\Lambda} is dominated by the cosmological
constant, as it in expected for the SM. In the symmetric phase
\mbo{V/V'>0} and hence \mbo{\phi_i > \phi_e\,.}  Note that a rescaling
of the potential does not affect inflation, but the relative weight of the
terms is crucial. In fact, SM Higgs inflation is far from being self-evident.
A detailed analysis is devoted to a forthcoming paper~\cite{NextPaper}.

For the SM Higgs
potential in the symmetric phase, denoting \mbo{z\equiv
\frac{\lambda}{6\,m^2}\,,} and a potential $V=V(0)+\Delta V(\phi)$ we have a term 
\mbo{\frac{V(0)}{2m^2}\frac{1}{\phi}\frac{1}{1+z\phi^2}} plus
\mbo{\frac{\Delta V}{V'}=\frac{\phi}{4}\,\left(1+\frac{1}{1+z\phi^2}\right)}
and thus with
\ba
{\cal I}=\int\limits_{\phi_e}^{\phi_i}\,\frac{V}{V'}\,\D \phi=
\frac{V(0)}{2m^2}\left[\ln \frac{\phi_i^2}{\phi_e^2}-\ln \frac{
 {\phi_i}^2\,z+1}{{\phi_e}^2\,z+1}\right]+ 
\frac18\,\left[\phi_i^2-\phi_e^2
+\frac{1}{z}\,\ln \frac{
 {\phi_i}^2\,z+1}{{\phi_e}^2\,z+1}\right]
\ea
we obtain
\bea
N_e=\frac{8\pi}{\mpl^2}\,{\cal I}\epo
\label{Neformula}
\eea
Note that $V(0)=\frac{m^2}{2}\,\Xi+\frac{\lambda}{8}\,\Xi^2$ with
$\Xi=\frac{\mpl^2}{16\pi^2}$, $m^2$ and
$z=\frac{\lambda}{6m^2}$ all are known SM quantities! 
\mbo{N_e} large requires \mbo{\phi_i \gg \phi_e}. In our calculation,
adopting \textbf{fixed} parameters as given at the Planck scale, and
$\phi_i\simeq 4.51 \mpl$ as an initial field one obtains $\phi_e=-1.32 \power{-6} \mpl$ which
yields $N_e\approx 65.83$. The Higgs field in this constant coupling
approximation starts oscillating for times $t\,\gapprox\, 200\,
\mpl^{-1}$. If we take into account the \textbf{running} of parameters as given
by the standard \MSb RG we find $\phi_e \simeq 3.73\power{-5} \mpl$
and $N_e \approx 65.05$ a value not too far above the
phenomenologically required minimum bound. The Higgs field in this
more adequate calculation is found to oscillate at much later times
but still before the Higgs transition (see Fig.~\ref{fig:Lagrangianext} below).
A detailed analysis shows that the dynamical part of the Higgs potential
$\Delta V(\phi)$ decays exponentially, while $V(0)$ the quasi
cosmological constant is weakly scale dependent through $m^2_0(\mu)$
and $\lambda(\mu)$, but has a zero not far above the Higgs transition
point, as can be seen in Fig.~\ref{fig:Lagrangianext} (see
Ref.~\cite{NextPaper} for details). 

The inflation scenario suggested by the present analysis is a Gaussian
potential with small anharmonic perturbations, since $m^2$ is
predicted to be large while $\lambda$ remains small. This picture
should be valid in the renormalizable effective field theory regime
below about $10^{17}~\gv$. Going to higher energies details of the
cutoff system are expected to come into play, effectively in form of
dimension 6 operators as leading corrections. These corrections are
expected to get relevant only closer to the Planck scale.
\begin{figure}
\centering
\includegraphics[height=6cm]{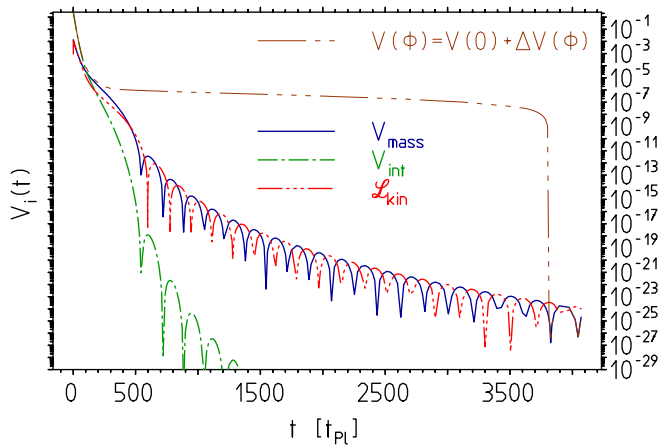}
\caption{The mass-, interaction- and kinetic-term of the bare
Lagrangian in units of $\mpl^4$ as a function of time. The vacuum term
$V(0)$ gets nullified across the vacuum rearrangement somewhat above
the Higgs transition point.}
\label{fig:Lagrangianext} 
\end{figure}
I expect that the observed value of dark energy has to be considered
as a phenomenological constraint. The reason is that $\rho_\Lambda$ is
dependent on the Higgs field magnitude, which is not fixed by other
observations, except maybe by CMB inflation data. In addition we have
to keep in mind that our scenario is very sensitive to the basic
parameters $C(\mu)$ and $\lambda(\mu)$, which were obtained by
evolving coupling parameters over 16 orders of magnitude in
scale. This cries for high precision physics to really settle the
issue. High precision physics could become the tool in probing and
investigating early cosmology.  Note that given the SM couplings
everything is essentially (besides the Higgs field strength) a SM
prediction without any extra assumption. What we also learn is that
the quartic ``divergences'' contributing to the vacuum energy, like
the quadratic ``divergences'' affecting the Higgs mass, play an
important role in promoting the SM Higgs to the inflaton, and
inflation to be an unavoidable consequence of the SM.

\section{Summary and Outlook}
We adopt that the new particle found by ATLAS and CMS at the LHC is
the SM Higgs and we argue about the specific value found for the Higgs
mass $M_H=125.9\pm 0.4~\gv$ and its impact for the SM itself. As noted
quite some time ago in Ref.~\cite{Hambye:1996wb} stability of the SM
vacuum up to the Planck scale is just what it now turns out to be and
this looks to be more than just an accident. It signals a higher
self-consistency of the SM than anticipated before. Provided the Higgs
potential remains stable, there is no non-perturbative Higgs issue, no
Landau pole nor any other problem. Surprisingly, except from the moderately
increasing Abelian $U(1)_Y$ gauge coupling, all other effective
couplings behave asymptotically free, which renders Planck scale
physics accessible by perturbative methods. The amazing thing is that
this is the result of the an intricate conspiracy of the several
interactions ``unified'' in the SM. Besides the Higgs self-coupling
$\lambda$, the top-quark Yukawa coupling $y_t$, the strong interaction
coupling $\alpha_s$, as well as the gauge couplings $g$ and $g'$ turn
out to be important in the conspiracy responsible for the stability of
the SM ground state. Note that the full knowledge of the RG
coefficients is needed to obtain a stable solution, while
approximations like the gaugeless one may suggest a false metastable
situation. I also would say that complete four-loop calculations of
the $\beta$-function coefficients would be highly desirable.

Concerning the vacuum stability, we should keep in mind: the Higgs
mass miraculously turns out to have a value as it was expected form
vacuum stability considerations, given the unexpectedly heavy
top-quark. As we have seen it is a tricky conspiracy among the SM
couplings which allows for a regular and stable extrapolation up to
the Planck scale. If the SM misses to have a stable vacuum, why does
it just miss it almost not as several related analyses
Refs.~\cite{Holthausen:2011aa,Yukawa:3,degrassi,Moch12,Mihaila:2012fm,Chetyrkin:2012rz,Masina:2012tz,Bednyakov:2012rb,Bednyakov:2012en,Bednyakov:2013eba,Chetyrkin:2013wya,Buttazzo:2013uya}
find? As also discussed in
Refs.~\cite{Bian:2013jra,Masina:2013wja,Hamada:2013mya}, very
different scenarios would follow if the main condition of vacuum
stability and the existence of a sign change of the Higgs potential
mass term below the Planck mass scale would not be satisfied. At
present the Higgs potential stability issue looks to be almost
entirely a matter of the precise value of the \MSb top-quark Yukawa
coupling at an appropriate matching scale. This issue is not settled
in my opinion. First of all we are left with the question what the top
quark mass measured by experiments precisely means. For our analysis
we have identified the PDG top-quark mass entry with the on-shell
mass. The second point concerns the missing higher order corrections
in the matching conditions, which could help to clarify the
situation. Problems in this direction have been discussed in
Ref.~\cite{Jegerlehner:2012kn} (see also Ref.~\cite{Buttazzo:2013uya}
and references therein).

If our LEESM scenario is realistic, meaning that there is no essential
non-SM physics up the Planck scale, the Higgs not only provides masses
to the SM particles, but also supplies the necessary dark energy
triggering inflaton.  To settle the issues of inflation, the EW phase
transition and baryogenesis, a very precise knowledge of the SM
parameters becomes more crucial than ever. To achieve much better
control on SM parameter-evolution, over 16 orders of magnitude in
scale, becomes a key issue of particle physics and early cosmology. If
we look at the leading coefficient of the quadratic divergence
(\ref{coefC1}) we see that the top Yukawa term is enhanced by a factor
6 relative to the Higgs coupling, which means that a precise top-quark
Yukawa coupling measurement is most crucial and should have highest
priority at an ILC, where a threshold scan could provide much more
reliable information (see
e.g.~\cite{Hoang:1998xf,Hoang:2000yr,Baer:2013cma,Agashe:2013hma}).
This is of course an issue only if there is not a lot of yet unknown
stuff which could obscure the situation. Almost equally important is a
precise knowledge of the Higgs self-coupling. Especially the inflation
data are constraining the possible values in $(\lambda,y_t)$-plane at
$\mpl$ dramatically. It is very surprising that such a possible window
actually seems to exist. It also implies that higher order
perturbative corrections are more important than ever, as a tool to
deepen our understanding of fundamental phenomena. Precision physics
maybe a key tool to monitor the unknown bare physics at very high scales.

We understand the SM as a low energy effective emergence of some
unknown physical system, we may call it ``ether'', which is residing
at the Planck scale with the Planck length as a ``microscopic'' length
scale. Note that the cutoff, though very large, in any case is
finite. Correspondingly, counterterms are finite. In such a kind of
low energy effective scenario quadratic and quartic ``divergences''
play an important role when we approach the bare system at the Planck
scale. One key quantity here is the Higgs mass counterterm, which is
given by $\delta M_H^2=\frac{M_{\rm Pl}^2}{16 \pi^2}\, C(\mu)$,
with $C(\mu=\mpl)\simeq 0.282$, where $C$ is the coefficient of the
quadratic divergence of the bare Higgs mass given in
Eqs.~(\ref{quadraic2},\ref{coefC2}). Note that the bare mass
$m(\mu=\mpl)/\mpl=0.0295$ is almost two orders of magnitude below
$\mpl$. Note that in the broken phase at the EW scale $C(\mu=v)\sim
-6.7$. Our main observation is that for appropriate input parameters
the quadratically enhanced Higgs mass counterterm as a function of the
renormalization scale exhibits a zero somewhat below the Planck
scale. The zero implies a change of sign of the bare Higgs mass, which
is responsible for the Higgs mechanism as a first order phase
transition at about $\mu_0\sim 1.4 \times 10^{16}~\gv$ in the $T=0$
SM. Above this scale the system is in the unbroken phase i.e. the
Higgs VEV is vanishing and all modes besides the remaining complex
Higgs doublet fields ideally are massless. The second quantity which
is dramatically enhanced by cutoff effects is the cosmological
constant $V(0)=\braket{V(\phi)}$ (Higgs vacuum loops), which yields a
contribution to the dark energy density $\delta
\rho_\Lambda=\frac{M_{\rm Pl}^4}{(16 \pi^2)^2}\, X(\mu)$, where
$X(\mu)=\frac18\, \left(2\,C(\mu)+\lambda(\mu)\right)$ with
$X(\mu=\mpl)\simeq 0.088$. As $\lambda(\mu_0)$ is small $X(\mu)$ has a
zero not far below the zero of $C(\mu)$. Thus, as for the Higgs mass,
there is a matching point between the renormalized low energy
cosmological constant and the bare one seen at short distances. Again,
that the bare cosmological constant is huge in the symmetric phase is
not in conflict with the observed tiny dark energy density of today.

In the symmetric phase we naturally have very heavy Higgses, while the
light Higgs in the broken phase is a consequence of the phase
transition itself, because all SM masses, including the Higgs itself,
are proportional to the Higgs VEV $v$, which is an order parameter and
hence naturally is a low energy quantity. The key point is that before
the Higgs mechanism has taken place the large positive bare Higgs
mass-square term in the Higgs potential provides a huge dark energy
term which triggers inflation and the four heavy Higgses represent the
inflaton. Slow-roll inflation ends because of the exponential decay of
the Higgs fluctuation fields after short time, while $V(0)$ persists to
be large until it is nullified somewhat before the Higgs transition point
and the subsequent EW phase transition.

The EW phase transition, due to finite temperature effects, takes
place always a little below the Higgs transition scale. In our case
the EW transition essentially coincides with the Higgs transition,
i.e. it takes place at $\mu_0\sim 1.4 \power{16}~\gv$ not near
$\mu\sim v \simeq 246.22~\gv$ or elsewhere far below a typical GUT
scale. This must have a definite impact on baryogenesis, and commonly
accepted assumptions (see
e.g. Refs.~\cite{Dine:1992wr,Buchmuller:1995sf,Kajantie:1996mn,Rubakov:1996vz,Chung:2012vg}
and references therein) have to be reconsidered.

In the symmetric phase, during inflation, the heavy Higgses decay into
massless fermions, which provides the reheating of the universe which
dramatically cools down by the inflation. The Higgs decay width
$\Gamma(H\to \bar{f}f)=\frac{M_H\,y_f^2}{16\pi}$ can be large for
massless fermions, depending on the Yukawa couplings. Produced are
preferably the modes with largest Yukawa coupling like the yet
massless $t\bar{t}$- $t\bar{b}$- $b\bar{t}$-pairs, the latter two
modes via the ``charged'' Higgses and rates proportional to
$y_ty_b$. While $b\bar{b}$-production is suppressed by
$y^2_b/y^2_t(\mpl)\sim4.4\power{-4}$, $\tau$-production follows with a
branching fraction $y^2_\tau/(3\,y^2_t)(\mpl)\sim2.4\power{-4}$
etc. During and/or shortly after the EW phase transition, the heavier
quarks decay into the lighter ones (the strongly coupled into the
weakly coupled) channeled by the CKM-matrix~\cite{CKM}. Therefore most of the
normal matter is a decay product of top and bottom quarks and
their anti-quarks. In this scenario most normal matter must have undergone
CP-violating decay processes. This is certainly an important
ingredient for baryogenesis.

Here we also should remind that QED, the electric charge assignments
and massless photon radiation etc are only defined after symmetry
breaking $SU(3)_c\otimes SU(2)_L\otimes U(1)_Y \to
SU(3)_c\otimes U(1)_{\rm em}$. This certainly requires
more detailed studies including the question which scalars couple to
the vacuum. Another interesting aspect: in the symmetric phase
$SU(2)_L$ is unbroken in the very early phase of the universe. In the
symmetric phase there could exist heavy $SU(2)$ bound states\footnote{Note that the
basic parameters of this $SU(2)$ is known and below close to the Planck scale actually
has a coupling slightly stronger than the one of the $SU(3)$ sector
(see Table \ref{tab:params}).} which
would bind energy and could be dark matter candidates. Thus cold dark
matter could be frozen energy, very similar to ordinary matter, which
is mainly hadronic binding energy (nucleon masses), while the masses
of the elementary fields induced by the Higgs mechanism constitute an
almost negligible contribution to normal matter in the universe. The
way matter clusters and populates the universe of course is determined
by the details of the Yukawa- and Higgs-sector and the particular form
of the EW phase transition.

A final remark about the reliability of numerical estimates given:
precise numbers are expected to change not only when input parameters
change, but may be affected by including higher order corrections. The
role of what matching conditions precisely are used and justified to
convert physical into \MSb parameters remains a key issue. One also
should be aware that what we used as a Planck scale in terms of the
\MSb scale $\mu$ may differ by non-trivial factors.  If the Planck medium
would be a lattice of spacing $\aPl$ the effective Planck energy scale
could be $\tilde{\Lambda}_{\rm Pl}=\pi/\aPl=\pi \mpl$ a factor of
$\pi$ higher that what we have assumed throughout the analysis.
A similar ambiguity may be due to the \textbf{convention} adopted when
defining the \MSb scheme. We may ask whether the minimal \MS parametrization
would not be more closely related to the bare parameters, relevant at the
Planck scale, than the \MSb ones.
We have
$\ln \mu^2_\mathrm{MS}=\ln \mu^2_{\MSbm}+\gamma-\ln 4 \pi$
thus
$\mu_{\MSbm}/\mu_\mathrm{MS}=\exp \frac{\ln 4 \pi -\gamma}{2}\simeq
2.66\,,$
which tells us that we should be aware of the fact that the
identification of the renormalization scale with a physical Planck scale may be
fairly arbitrary within some $O(1)$ factors. What it means is that the
matching scale looks to be ambiguous within a factor $\sim3$, while the
beta-functions do not change. While the \MSb
parametrization is fixed at the EW scale phenomenologically (up to
possible matching condition uncertainties) it is conceivable that
the identification $\mu=\mpl$ is requiring phenomenological adjustment
as well, via indirect constraints from properties of the EW phase transition and the
observed inflation profile, for example.

We have not discussed the possibility that the sterile right-handed
neutrinos, which must exist in order to allow for non-vanishing
neutrino masses, are Majorana particles. In this case, it would be
natural that the singlet Majorana neutrinos have huge masses in the
symmetric as well as in the broken phase, not protected by any of the
SM symmetries. As is well known, this would provide the most natural
explanation for the smallness of the neutrino masses by the resulting
seesaw mechanism. When the singlet neutrino mass term is very high
near $\mpl$ it would not affect the running of the other couplings
also because a singlet has not any direct couplings to other
fields. In addition, as the singlet Majorana mass is not subject to
mass-coupling relations (it is intrinsic and not generated by a Higgs
type mechanism) it is actually decoupling, although it leaves its trace
in scaling the neutrino masses to very small values. Nevertheless, it
would be interesting to investigate the scale dependence of the
effective heavy Majorana masses and to study their influence on
inflation, the EW phase transition and the dark matter problem, within
the LEESM framework.

We also remark that gravity as we see it at long distances in our
scenario emerges form the ``ether'' system exhibiting an intrinsic
fundamental cutoff. Therefore, there is no reason why we should expect
gravity to be quantized in the sense of e local relativistic
renormalizable QFT at the Planck scale. Also for gravity the low
energy manifestation is expected to be what is obtained form a low
energy expansion (see Ref.~\cite{Jegerlehner:1978nk} for a
corresponding consideration).

Our findings do not exclude the existence of new physics as far as it
does not spoil the gross features of the SM which are important: the
stability of the Higgs potential and the existence of a zero in the
coefficient of the quadratic divergences. Also important for the
understanding of the today very small dark energy density is the zero in
the coefficient of the corresponding quartically enhanced contribution.

In any case after all relevant ingredients of the SM are confirmed and
parameters have been determined within narrow error bands many issues
in early cosmology likely are direct predictable consequences of
properties of the Higgs system and its embedding into the SM. This
insight opens completely new possibilities for the solution of open
problems.  So far the LEESM scenario has more phenomenological support
than any of the other known beyond the SM scenarios.  However, the
inflation scenario is very sensitive to the precise SM parameter input
values.  This is not surprising as we try to extrapolate over 16
orders of magnitude in scale. In addition, there are two unknown
inputs in the game which affect inflation. One is the magnitude of the
Higgs field near the Planck scale, the other the renormalized mass $m$
in the symmetric phase. These two inputs are constrained by what we
think to know about inflation, slow-roll, equation of state,
Gaussianity, spectral index, in particular. After all the SM hides
more secrets than answers and we are far from having worked out all
its consequences nor have we understood many of its \textit{why
so}'s. This is an attempt to understand the SM as a conspiracy
theory. More and more the SM looks to me to work like a fine Swiss
clockwork.

As a final remark about the role of the SM Higgs let me point out the
following: for some time at and after the big bang the Higgs is the
particle which is directly attached to gravity. It is the only SM
particle which directly talks to the vacuum in the early universe
(much later QCD phase transition also quark and gluon condensates).
The Higgs is the one producing negative pressure and hence blowing
continuously energy into the expanding universe. Amazingly,
understanding the physics of the early universe now depends vitally on
the precise determination and understanding of parameters like the
top quark mass and the Higgs mass, and the precise values of their
couplings. Seeing more of the ``ether'' residing at the Planck scale
is now a matter of high precision physics. 

The key questions are ``Where in SM or SM+ parameter space is the hot
spot, which makes the Higgs be the inflaton?'' ``Does the Higgs play
the master role in the early evolution of our universe?'' Higher order
effects, moderate additions to the SM like a
Peccei-Quinn~\cite{Peccei:1977hh} axion sector and its impact on the
strong CP problem could also still play a role in the fine-tuning
conspiracy. On the other hand what looks to be a straight forward
possible renormalizable extension of the SM like a fourth fermion
family seem definitely ruled out in our scenario.

In any case our LEESM Higgs scenario offers a number of new aspects not
considered so far and are worth being investigated in much more
detail.  Key point of the present analysis are the \MSb input
parameters at the $Z$ mass scale, evaluated in terms of the matching
conditions as studied in Ref.~\cite{Jegerlehner:2012kn} for central
values $M_H=126~\gv$ and $M_t=173.5~\gv$. Any kind of possible and
necessary fine tuning, by adjusting $\lambda(M_H)$ or $y_t(M_t)$, has
not been analyzed in detail so far. Moderate tuning or more accurate
predictions of the SM input certainly will provide a more reliable
prediction of the SM Higgs inflation pattern. The sensitivity to
details is pronounced. The more it is remarkable that we found the
spot where the SM provides dark energy and inflation
``automatically'', and reasonably close to what we know from
observation. For the first time we have a chance to get information
about inflation from the SM alone and we can make predictions which
are not just more or less direct consequences of some more of less
plausible assumptions. The main new point is that we find the Higgs
potential to be stable up to the Planck scale and that the coefficient
of the $\mpl^2$ enhanced Higgs potential mass term changes sign
sufficiently below the Planck scale.
	
\bigskip

\noindent
\newpage

\section*{Acknowledgments}

I am grateful to Mikhail Kalmykov for many inspiring discussions, for
critically reading the manuscript as well as for long-time
collaboration in electroweak two-loop calculations and in particular
on working out the relationship between on-shell and \MSb parameters
in the SM, which play a key role in the present work. I would like to
thank Oliver B\"ar, Nigel Glover and Daniel Wyler for their interest
and for many clarifying discussions.  I also thank for support by the
EC Program {\it Transnational Access to Research Infrastructure}
(TARI) INFN - LNF, HadronPhysics3 - Integrating Activity, Contract
No.~283286.

\appendix

\section*{Appendix: the ingredients for one-loop matching}
The one-loop on-shell counterterms may be expressed in terms of the
known scalar integrals:
\begin{eqnarray*}
A_0 (m) &=& - m^2 (\Reg + 1 - \ln m^2)\\
B_0 (m_1, m_2; s) &=& \Reg - \int ^1 _0 \D z \ln ( - sz ( 1 - z) +
m^2_1 (1 -z) + m^2_2 z-\I \veps)\;,
\end{eqnarray*}
with
\begin{eqnarray*}
\Reg = \frac{2 }{ \eps} - \gamma + \ln 4 \pi + \ln \mu _0^2 \equiv \ln
\mu^2 \epo
\end{eqnarray*}
In addition we define $\cC=\frac{\sqrt{2}\,G_\mu}{16\,\pi^2}$
and $\cC_\mu=\frac{\sqrt{2}\,G_\mu}{16\,\pi^2}\, \ln
\mu^2$. Furthermore, $c_W^2=\frac{M_W^2}{M_Z^2}$ and
$s_W^2=1-c_W^2$. Sums over fermion contributions we write $\sum_{f_s}$
for sums over individual fermions, and $\sum_{f_d}$ for sums over
fermion doublets. $Q_f$ denotes the fermion charge, $a_f=Q_f\,s_W^2\mp\frac14$ the
$Zf\bar{f}$ vector coupling and by $b_f=\pm \frac14$ the axial-vector couplings.
A color factor $N_c=3$ applies for quarks.

\MSb counterterms:
{\small
\bea
\left.\frac{\delta v^{-1}}{v^{-1}}\right|_{\MSbm}  &=&
  + \cC_\mu \, ( 3\,\frac{M_Z^4}{M_H^2} + 6\,\frac{M_W^4}{M_H^2} - \frac32\,M_Z^2
  - 3\,M_W^2 + \frac32\,M_H^2 \crn &&+ \sum_{f_d} [-4\,\frac{m_1^2+m_2^2}{M_H^2}+m_1^2+m_2^2])\crn
\left.\frac{\delta M_Z^2}{M_Z^2}\right|_{\MSbm} &=&
  + \cC_\mu \, (  - 6\,\frac{M_Z^4}{M_H^2} - 12\,\frac{M_W^4}{M_H^2} + \frac{11}{3}\,M_Z^2
  + \frac{14}{3}\,M_W^2 - 28\,c_W^2\,M_W^2 - 3\,M_H^2 \crn &&+ \sum_{f_s} [8\,\frac{m_f^4}{M_H^2}-2\,m_f^2+\frac{22}{27}\,
(M_Z^2-2\,M_W^2)+\frac{40}{27}\,c_W^2M_W^2])\crn
\left.\frac{\delta M_W^2}{M_W^2}\right|_{\MSbm} &=&
  + \cC_\mu \, (  - 6\,\frac{M_Z^4}{M_H^2} - 12\,\frac{M_W^4}{M_H^2} + 3\,M_Z^2
  - \frac{68}{3}\,M_W^2 - 3\,M_H^2 \crn &&+ \sum_{f_d} [8\,
    \frac{m_1^4+m_2^4}{M_H^2}-2\,(m_1^2+m_2^2)+\frac43\, M_W^2] )\crn
\left.\frac{\delta M_H^2}{M_H^2}\right|_{\MSbm} &=&
  + \cC_\mu \, (  - 3\,M_Z^2 - 6\,M_W^2 + 3\,M_H^2 + \sum_{f_s} [2\,m_f^2])\crn
\left.\frac{\delta m_t}{m_t}\right|_{\MSbm} &=&
  + \cC_\mu \, ( 12\,\frac{M_t^4}{M_H^2} + 12\,\frac{M_b^4}{M_H^2}
    - 3\,\frac{M_Z^4}{M_H^2} - 6\,\frac{M_W^4}{M_H^2}
    - \frac32\,M_b^2 + \frac32\,M_t^2 - \frac43\,M_Z^2 \crn &&\hspace*{10mm}
    + \frac{20}{3}\,M_W^2 - \frac{16}{3}\,c_W^2\,M_W^2 -
      \frac32\,M_H^2
    - \frac{16}{3}\,c_W^2\,s_W^2\,M_Z^2)\crn
\left.\frac{\delta m_b}{m_b}\right|_{\MSbm} &=&
  + \cC_\mu \, ( 12\,\frac{M_t^4}{M_H^2} + 12\,\frac{M_b^4}{M_H^2}
     - 3\,\frac{M_Z^4}{M_H^2} - 6\,\frac{M_W^4}{M_H^2}
     + \frac32\,M_b^2 - \frac32\,M_t^2 + \frac23\,M_Z^2 \crn && \hspace*{10mm}
     + \frac23\,M_W^2 - \frac43\,c_W^2\,M_W^2 - \frac32\,M_H^2
     - \frac43\,c_W^2\,s_W^2\,M_Z^2) \nn
\eea
}
On-shell counterterms:
{\small
\bea
\frac{\delta e}{e} &=& \cC \,s_W^2M_W^2\,\Big(\frac{38}{3}+14\,\frac{A_0(M_W)}{M_W^2}
- \frac83\,\sum_{f_s}Q_f^2 \,\big(1+\frac{A_0(m_f)}{m_f^2}\big)\Big)\crn
\frac{\delta v^{-1}}{v^{-1}}&=& \frac{\delta e}{e}
     -\frac{1}{2\,s_W^2}\,\big(s_W^2\,\frac{\delta
     M_W^2}{M_W^2}+c_W^2\,\frac{\delta M_Z^2}{M_Z^2}\big) \nn 
\eea
\bea
\delta M_H^2&=&\cC\,\Big(
      A_0(M_H)       \,(3\,M_H^2)\crn&&
     +A_0(M_Z)       \,(M_H^2+6\,M_Z^2)\crn&&
     +B_0(M_H,M_H,M_H^2)\,(\frac92\,M_H^4)\crn&&
     +B_0(M_Z,M_Z,M_H^2)\,(\frac12\,M_H^4-2\,M_H^2\,M_Z^2+6\,M_Z^4)\crn&&
     +A_0(M_W)       \,(2\,M_H^2+12\,M_W^2)\crn&&
     +B_0(M_W,M_W,M_H^2)\,(M_H^4-4\,M_H^2\,M_W^2+12\,M_W^4)\crn&+&\sum_{f_s}\big[
 A_0(m_f)\,(-8\,m_f^2)+B_0(m_f,m_f,M_H^2)\,(2\,M_H^2\,m_f^2-8\,m_f^4)\big]\Big)\nn
\eea
\bea
\delta M_Z^2&=&\cC\,\Big(
     -\frac23\,M_H^2\,M_Z^2+4\,\frac{M_Z^4}{M_H^2}\,M_Z^2-\frac29\,M_Z^4\crn&&
     +8\,\frac{M_W^4}{M_H^2}\,M_Z^2-\frac{20}{9}\,M_W^2\,M_Z^2+\frac{16}{3}\,M_W^4-16\,c_W^2\,M_W^4\crn&&
     +A_0(M_H)       \,(\frac13\,M_H^2+2\,M_Z^2)\crn&&
     +A_0(M_Z)       \,(-\frac13\,M_H^2+6\,\frac{M_Z^4}{M_H^2}+\frac23\,M_Z^2)\crn&&
     +A_0(M_W)       \,(12\,\frac{M_W^2}{M_H^2}\,M_Z^2+\frac{16}{3}\,M_W^2-16\,c_W^2\,M_W^2+\frac23\,M_Z^2)\crn&&
     +B_0(M_Z,M_H,M_Z^2)\,(-\frac43\,M_H^2\,M_Z^2+\frac13\,M_H^4+4\,M_Z^4)\crn&&
     +B_0(M_W,M_W,M_Z^2)\,(\frac{16}{3}\,M_W^2\,M_Z^2-\frac{68}{3}\,M_W^4-16\,c_W^2\,M_W^4+\frac13\,M_Z^4)\crn
&+&\sum_{f_s}\big[
(\frac{32}{3}\,M_Z^2\,m_f^2-\frac{16}{9}\,M_Z^4)\,(a_f^2+b_f^2)\crn &+&
  A_0(m_f)\,(-8\,\frac{M_Z^2}{M_H^2}\,m_f^2+\frac{32}{3}\,M_Z^2\,(a_f^2+b_f^2))\crn &+&
  B_0(m_f,m_f,M_Z^2)\,(\frac{32}{3}\,M_Z^2\,m_f^2\,(a_f^2-2b_f^2)
+\frac{16}{3}\,M_Z^4\,(a_f^2+b_f^2))\big]\Big) \nn 
\eea
\bea
\delta M_W^2&=&\cC\,\Big(
     -\frac23\,M_H^2M_W^2+8\,\frac{M_W^6}{M_H^2}+4\,\frac{M_Z^4}{M_H^2}M_W^2-\frac{112}{9}\,M_W^4-\frac23\,M_Z^2M_W^2\crn&&
     +A_0(M_H)       \,(2\,M_W^2+\frac13\,M_H^2)\crn&&
     +A_0(M_Z)       \,(6\,\frac{M_Z^2}{M_H^2}M_W^2+\frac83\,M_W^2-8\,c_W^2M_W^2+\frac13\,M_Z^2)\crn&&
     +A_0(M_W)       \,(12\,\frac{M_W^4}{M_H^2}-4\,M_W^2-\frac13\,M_H^2-\frac13\,M_Z^2)\crn&&
     +B_0(M_H,M_W,M_W^2)\,(-\frac43\,M_H^2M_W^2+\frac13\,M_H^4+4\,M_W^4)\crn&&
     +B_0(M_Z,M_W,M_W^2)\,(-\frac{68}{3}\,M_W^4-16\,c_W^2M_W^4+\frac13\,M_Z^4+\frac{16}{3}\,M_Z^2M_W^2)\crn&&
     +B_0(0,M_W,M_W^2)\,(-16\,s_W^2M_W^4)\crn&+&\sum_{f_d}\big[
\frac43\,M_W^2\,(m_1^2+m_2^2)-\frac49\,M_W^4\crn &+&
  A_0(m_1)\,(-8\,\frac{M_W^2}{M_H^2}\,m_1^2+\frac43\,M_W^2-\frac23\,(m_1^2-m_2
^2))\crn &+&
  A_0(m_2)\,(-8\,\frac{M_W^2}{M_H^2}\,m_2^2+\frac43\,M_W^2+\frac23\,(m_1^2-m_2
^2))\crn &+&
  B_0(m_1,m_2,M_W^2)\,(\frac43\,M_W^4-\frac23\,M_W^2\,(m_1^2+m_2^2)
                             -\frac23\,(m_1^2-m_2^2)^2) \big] \Big)\nn
\eea
\bea
\delta m_\tau&=&m_\tau\, \cC\,\Big(
      4\,\frac{M_W^4}{M_H^2}+2\,\frac{M_Z^4}{M_H^2}-3\,M_W^2+\frac32\,M_Z^2\crn&&
     +A_0(M_H)      \crn&&
     +A_0(M_Z)       \,(3\,\frac{M_Z^2}{M_H^2}+6\,\frac{M_W^2}{m_\tau^2}-4\,c_W^2\frac{M_W^2}{m_\tau^2}-\frac52\,\frac{M_Z^2}{m_\tau^2})\crn&&
     +A_0(M_W)       \,(6\,\frac{M_W^2}{M_H^2}-\frac{M_W^2}{m_\tau^2}+\frac12)\crn&&
     +A_0(m_\tau)       \,(-2\,\frac{M_W^2}{m_\tau^2}+\frac52\,\frac{M_Z^2}{m_\tau^2}+1)\crn&&
     +B_0(m_\tau,M_H,m_\tau^2)\,(-\frac12\,M_H^2+2\,m_\tau^2)\crn&&
     +B_0(m_\tau,M_Z,m_\tau^2)\,(6\,M_W^2\,M_Z^2/m_\tau^2+12\,M_W^2-8\,c_W^2M_W^2-4\,\frac{M_W^4}{m_\tau^2}-\frac72\,M_Z^2-\frac52\,\frac{M_Z^4}{m_\tau^2})\crn&&
     +B_0(m_\tau,0,m_\tau^2)\,(-8\,M_W^2+8\,c_W^2M_W^2)\crn&&
     +B_0(m_{\nu_\tau},M_W,m_\tau^2)\,(\frac12\,M_W^2-\frac{M_W^4}{m_\tau^2}+\frac12\,m_\tau^2)+
\sum_{f_s} A_0(m_f) \, ( - 4\,\frac{m_f^2}{M_H^2})\Big)\nn
\eea
\bea
\delta M_b &=&M_b\,\cC\,\Big( - \frac49\,s_W^2M_W^2
 + 2\,\frac{M_Z^4}{M_H^2} + 4\,\frac{M_W^4}{M_H^2} - \frac{13}{18}\,M_Z^2 - \frac{11}{9}\,M_W^2 + \frac49\,\frac{M_W^4}{M_Z^2} \crn&&
 + A_0(M_H) \, \crn&&
 + A_0(M_W) \, ( \frac12 + 6\,\frac{M_W^2}{M_H^2} - \frac12\,\frac{M_t^2}{M_b^2} - \frac{M_W^2}{M_b^2} )\crn&&
 + A_0(M_Z) \, ( 3\,\frac{M_Z^2}{M_H^2} - \frac{5}{18}\,\frac{M_Z^2}{M_b^2} + \frac29\,\frac{M_W^2}{M_b^2} - \frac49\,\frac{M_W^2}{M_Z^2}\frac{M_W^2}{M_b^2} )\crn&&
 + A_0(M_t) \, ( \frac12 + \frac12\,\frac{M_t^2}{M_b^2}+ \frac{M_W^2}{M_b^2} )\crn&&
 + A_0(M_b)\,4\,s_W^2 \, ( \frac13\,\frac{M_W^2}{M_b^2} )\crn&&
 + A_0(M_b) \, ( 1 + \frac{5}{18}\,\frac{M_Z^2}{M_b^2} - \frac29\,\frac{M_W^2}{M_b^2} + \frac49\,\frac{M_W^2}{M_Z^2}\frac{M_W^2}{M_b^2} )\crn&&
 + B_0(M_t,M_W,M_b^2) \, ( \frac12\,M_b^2 - M_t^2 + \frac12\,\frac{M_t^4}{M_b^2} + \frac12\,M_W^2 + \frac12\,M_W^2\,\frac{M_t^2}{M_b^2} - \frac{M_W^4}{M_b^2} )\crn&&
 + B_0(M_b,M_H,M_b^2) \, ( 2\,M_b^2 - \frac12\,M_H^2 )\crn&&
 + B_0(M_b,M_Z,M_b^2) \, ( \frac{17}{18}\,M_Z^2 - \frac{5}{18}\,\frac{M_Z^4}{M_b^2} + \frac49\,M_W^2 + \frac29\,M_W^2\,\frac{M_Z^2}{M_b^2} - \frac89\,\frac{M_W^4}{M_Z^2} - \frac49\,\frac{M_W^4}{M_b^2} )\crn&&
\sum_{f_s} A_0(m_f) \, ( - 4\,\frac{m_f^2}{M_H^2})\Big)\crn
\delta M_t &=&M_t\,\cC\,\Big(- \frac{16}{9}\,s_W^2M_W^2
 + 2\,\frac{M_Z^4}{M_H^2} + 4\,\frac{M_W^4}{M_H^2} - \frac{1}{18}\,M_Z^2 - \frac{29}{9}\,M_W^2 + \frac{16}{9}\,\frac{M_W^4}{M_Z^2}\crn&&
 + A_0(M_H) \, \crn&&
 + A_0(M_W) \, ( \frac12 + 6\,\frac{M_W^2}{M_H^2} - \frac12\,\frac{M_b^2}{M_t^2} - \frac{M_W^2}{M_t^2} )\crn&&
 + A_0(M_Z) \, ( 3\,\frac{M_Z^2}{M_H^2} - \frac{17}{18}\,\frac{M_Z^2}{M_t^2} + \frac{20}{9}\,\frac{M_W^2}{M_t^2} - \frac{16}{9}\,\frac{M_W^2}{M_Z^2}\frac{M_W^2}{M_t^2} )\crn&&
 + A_0(M_t)\,4\,s_W^2 \, ( \frac43 \frac{M_W^2}{M_t^2} )\crn&&
 + A_0(M_t) \, ( 1 + \frac{17}{18}\,\frac{M_Z^2}{M_t^2} - \frac{20}{9}\,\frac{M_W^2}{M_t^2} + \frac{16}{9}\,\frac{M_W^2}{M_Z^2}\frac{M_W^2}{M_t^2} )\crn&&
 + A_0(M_b) \, ( \frac12 + \frac12\,\frac{M_b^2}{M_t^2}+ \frac{M_W^2}{M_t^2} )\crn&&
 + B_0(M_t,M_H,M_t^2) \, ( 2\,M_t^2 - \frac12\,M_H^2 )\crn&&
 + B_0(M_t,M_Z,M_t^2) \, (  - \frac{7}{18}\,M_Z^2 - \frac{17}{18}\,\frac{M_Z^4}{M_t^2} + \frac{40}{9}\,M_W^2 + \frac{20}{9}\,M_W^2\,\frac{M_Z^2}{M_t^2} - \frac{32}{9}\,\frac{M_W^4}{M_Z^2} - \frac{16}{9}\,\frac{M_W^4}{M_t^2} )\crn&&
 + B_0(M_b,M_W,M_t^2) \, ( \frac12\,\frac{M_b^4}{M_t^2} -M_b^2 + \frac12\,M_t^2 + \frac12\,\frac{M_W^2}{M_t^2}\,M_b^2 + \frac12\,M_W^2 - \frac{M_W^4}{M_t^2} )\crn&&
\sum_{f_s} A_0(m_f) \, ( - 4\,\frac{m_f^2}{M_H^2})\Big)\nn
\eea
}
The on-loop corrections give the dominant contribution in the matching
relations. Two-loop
results may be found in Ref.~\cite{Jegerlehner:2001fb,Jegerlehner:2002er,Jegerlehner:2002em} and in Refs. quoted in Section~3.


\end{document}